\documentclass{article}[12pt]
\usepackage{graphicx} 
\usepackage{setspace} 
\usepackage{fullpage, epsfig, subfigure, wrapfig}
\usepackage{amssymb,amsmath}
\usepackage{tikz}
\usepackage[numbers]{natbib}

\newtheorem{definition}{Definition}{\bfseries}{\rmfamily}
\usetikzlibrary{positioning}

\usepackage{comment} 
\begin{document}


\title{Game Theory Meets LLM and Agentic AI: Reimagining Cybersecurity for the Age of Intelligent Threats}
\author{Quanyan Zhu}
\maketitle
%
%
\maketitle

\onehalfspacing
\abstract{
Protecting cyberspace requires not only innovative tools but also a foundational shift in how we reason about threats, trust, and autonomy. Traditional cybersecurity approaches rely heavily on manual effort, reactive responses, and brittle heuristics. To move toward proactive, trustworthy, and intelligent cyber defense, we need both new theoretical frameworks and software tools that work in concert rather than in parallel. On one end, game theory provides a rigorous foundation for modeling adversarial interactions, designing strategic defenses, and reasoning about trust in autonomous systems. It enables the development of frameworks that are proactive, adaptive, and grounded in rational decision-making. On the other end, software tools are being developed to process vast streams of cyber-relevant information, visualize attack surfaces, verify compliance, and recommend mitigation policies. While both efforts aim at securing digital systems, a disconnect remains between theory and practice.

The recent emergence of Large Language Models (LLMs) and agentic AI offers a transformative opportunity to bridge this gap. LLM-driven agentic AI can serve as a technological intermediary, translating abstract theoretical constructs into actionable, automated responses in real-world environments. Game theory can in turn inform the reasoning processes of agentic AI, guiding their decision-making and coordination across complex workflows. Moreover, LLM-powered agents are not merely implementers of game-theoretic ideas; they can also revolutionize game theory itself. Classical assumptions such as full rationality, perfect knowledge, and static payoff structures are increasingly untenable in modern cyber contexts. Agentic AI opens the door to rethinking these assumptions and building new game-theoretic models that better reflect the cognitive and computational realities of intelligent agents. This co-evolution promises novel solution concepts, new modeling paradigms, and richer theoretical foundations. From a systems engineering perspective, agentic AI also reshapes how we design software. It encourages modular, interactive, and continuously learning architectures. In parallel, concerns about trust, reliability, and security of AI agents must be addressed in the design process; these become first-class design principles rather than afterthoughts. 

This chapter introduces the foundations and convergence of game theory, LLM-based agentic AI, and cybersecurity. It is structured as follows. We begin by presenting game-theoretic modeling as a powerful approach for security analysis and defense strategy. Key frameworks—such as static games, dynamic games, signaling games, and Bayesian games—are reviewed, along with solution concepts including Nash equilibrium, Stackelberg strategies, and mechanism design, to show how they capture core aspects of cyber conflicts. We then shift focus to the capabilities of LLM-powered agents and their potential to address cybersecurity challenges. Next, we introduce LLM-driven game models that embed reasoning into AI agents and explore new LLM-based solution concepts. These models aim to advance the theoretical foundations of game theory in the LLM era and to establish quantitative metrics and design principles for deploying LLM agents in security-critical environments. Finally, we examine multi-agent workflows and the underlying games that govern their coordination. This chapter provides a comprehensive understanding of the interplay between game theory, agentic AI, and cybersecurity; and how their convergence can drive the development of secure, intelligent, and adaptive cyber systems.
}

\section{Introduction}
  
Cybersecurity is best understood as an environment shaped by the complex interactions among multiple agents: defenders, attackers, and users. Viewing the cybersecurity landscape through an agent-based lens helps capture the strategic, dynamic, and often asymmetric nature of these interactions. The complexity of cybersecurity has increased significantly over time. In the past, cyber threats were often characterized by one-shot interactions, where an attacker executed a single action to compromise a system. Today, threats are frequently multi-round in nature, as seen in Advanced Persistent Threats (APTs), where attackers engage in prolonged, adaptive campaigns that evolve over time.

Moreover, attackers are becoming increasingly intelligent and strategic. They are capable of reasoning, adapting to defenses, and using deception to mask their intentions or manipulate the defender’s perception of the state of the system. Attackers may possess information about system vulnerabilities unknown to defenders, creating a significant asymmetry in knowledge and initiative. Complicating matters further is the involvement of human users in these interactions. Users introduce additional layers of unpredictability, whether through misconfigurations, susceptibility to social engineering, or unintentional violations of security policies. These human factors not only increase the surface area of potential attack vectors but also make the design of robust and adaptive defense strategies more challenging.

Capturing the interactions among cybersecurity agents in a formal and systematic way is essential for understanding, modeling, and ultimately mitigating complex threats. There are two complementary approaches to formalizing these interactions: a systems-level modeling approach and a software/operational-level approach. The first approach involves systems-level representations of agent interactions. These include high-level graphical models, such as attack graphs, Bayesian networks, and influence diagrams. These tools are useful for abstracting and visualizing the structure of interactions among attackers, defenders, and users, and for reasoning about potential attack paths, dependencies, and probabilities of compromise. More fine-grained representations can incorporate additional dimensions, such as agents’ knowledge, capabilities, incentives, and intent, enabling richer models that bridge technical vulnerabilities with strategic behavior.

The second approach is software- or operation-level modeling, which focuses on capturing interactions at the implementation level. This includes monitoring function calls, control and data flows, system logs, and inter-process communications. By instrumenting and analyzing the behavior of actual software systems in real time, this approach enables detailed observation of agent actions and system responses, thereby supporting dynamic threat detection and response mechanisms. It also serves as a foundation for building adaptive tools that interact directly with the software environment to detect, respond to, or even anticipate attacks.

However, a critical gap exists between these two approaches. Systems-level models provide high-level abstractions and reasoning frameworks, but often lack the granularity and executability needed to inform real-time defense tactics. In contrast, software-level models offer operational precision and observability, but may lack the strategic context necessary to prioritize actions or anticipate adversarial intent. Bridging this gap requires aligning abstract representations of agent behavior and interactions with their concrete realizations in software systems. The system-level approach should guide the development and orchestration of software components by specifying expected behaviors, roles, and reasoning processes. Meanwhile, the software-level implementation should operationalize these specifications by embedding them into real-time execution environments. 

Modern agentic AI solutions are designed to perceive the environment, model adversarial strategies, infer user intent, and act, either by adapting system policies, deploying countermeasures, or simulating adversary behavior to improve resilience. By integrating strategic modeling with real-time system observability, agentic AI can enable proactive and intelligent defense mechanisms tailored to dynamic threat environments. A core objective is to create autonomous AI agents, software entities capable of interacting with other agents within a networked system, whether they are benign users, defenders, or adversaries. These agents must reason about their environment, anticipate adversarial moves, and coordinate defense actions, often under uncertainty and time pressure.

Game-theoretic system-level modeling tools provide the formal scaffolding for designing such agents. They specify the structure of interactions, encode incentives, capture asymmetries of information, and offer equilibrium concepts that guide the strategic behavior of agents. These models help define how an agent should reason about others’ intentions, adapt its strategy, and choose optimal responses in adversarial settings. At the same time, LLMs play a critical role in bridging the gap between abstract system-level models and concrete software implementations. LLMs can help generate code, translate high-level specifications into executable components, and synthesize policies or behaviors consistent with game-theoretic reasoning. Furthermore, LLMs can be embedded into agents themselves to enhance natural language understanding, support decision-making in open-ended environments, and facilitate coordination among heterogeneous agents. This integration of game-theoretic modeling and LLM-powered software synthesis lays the foundation for developing robust, adaptive, and interactive agentic AI systems that are both strategically sound and operationally viable within the evolving cybersecurity landscape.

This chapter is organized as follows. Section 2 introduces the role of game-theoretic modeling in cybersecurity, presenting foundational frameworks along with classical equilibrium concepts. It also outlines several baseline security games and explores their applications in cyber deception, network defense, and socio-economic cybersecurity challenges. Section 3 discusses the role of LLM  agents in cybersecurity and how they extend traditional game-theoretic approaches through dynamic reasoning and policy generation. Section 4 develops LLM-driven game models, focusing on two baseline formulations: the LLM-based Nash game and the LLM-based Stackelberg game. Section 5 examines multi-agent systems enabled by LLMs, where agent interactions are governed by local LLM-based game models. At the workflow level, the MAS architecture is governed by a gestalt game model that composes and coordinates multiple local interactions into a coherent global strategy. Section 6 concludes the chapter.

\section{Game-Theoretic Modeling}

Game theory plays a fundamental role in capturing and analyzing the complex interactions among defenders, attackers, and users in cybersecurity environments \cite{manshaei2013game}. These interactions are inherently strategic in nature, where each agent makes decisions based on their objectives, information, and predictions about how others will behave. Understanding these interdependent decisions is crucial for anticipating threats, optimizing defenses, and shaping user behavior.

In a typical cybersecurity scenario, the interaction space is defined by several critical dimensions. First, the agents or players involved include defenders (such as human operators, automated security systems, or network administrators), attackers (which may range from individual hackers to coordinated adversaries employing advanced persistent threats), and users (who might behave benignly, negligently, or maliciously if compromised). Each agent operates with a distinct set of information, often asymmetric. For instance, attackers may possess knowledge about undiscovered vulnerabilities, while defenders may have partial visibility into the system’s state. The capabilities of each agent, their available actions, also vary. Attackers might perform reconnaissance, escalate privileges, or exfiltrate data, while defenders may monitor network activity, deploy patches, reconfigure firewalls, or engage in deception. The objectives are likewise heterogeneous: attackers may seek disruption, data theft, or persistence, whereas defenders aim to minimize damage and maintain system integrity. Users typically prioritize performance, privacy, or usability, but their actions can inadvertently create opportunities for exploitation.

These interactions are rarely one-shot. Instead, they unfold over time as dynamic, multi-round engagements, especially in contexts involving stealthy and adaptive adversaries like APTs. The cybersecurity environment is often non-stationary, which means that the rules of interaction, available actions, or strategic objectives may evolve. New vulnerabilities may be discovered, defenses may be updated, and user behavior may change, creating a landscape that is constantly changing and requires continuous strategic adaptation.

To analyze such settings, game-theoretic models offer a range of frameworks depending on the nature of the interaction. One such class is the Stackelberg game, which models hierarchical or sequential interactions. In cybersecurity, this is especially relevant when defenders can commit to a strategy, such as deploying a specific defense configuration, before the attacker acts. The attacker then responds optimally given the observable commitment of the defender. These models are particularly useful in proactive defense planning, resource allocation, and deception strategies, where anticipating adversarial responses to defensive moves is crucial.

Another fundamental concept is the Nash game, in which agents act simultaneously without knowing the choices of the other. These models are applicable in decentralized environments, such as distributed systems or peer-to-peer networks \cite{fung2011smurfen,zhu2011game,zhu2012guidex}, where multiple agents make decisions independently. A Nash equilibrium represents a configuration in which no agent can improve their outcome by unilaterally changing their strategy. In cybersecurity, such models can be used to study competition over shared resources, insider threats, or collaborative defense settings.

To capture the temporal dimension and stochastic evolution of cyber-operations, Markov games, also known as stochastic games, are highly relevant. In these models, the system evolves through a sequence of states influenced by the joint actions of all agents. Each agent receives feedback in the form of rewards or penalties and updates its strategy over time. Markov games allow for long-term planning, reinforcement learning, and the modeling of state-dependent behavior. They are particularly suited to scenarios such as intrusion progression, escalation \cite{zhu2018multi,huang2020dynamic}, dynamic reconfiguration of defenses \cite{zhu2009dynamic,zhu2010network}, and adversarial learning \cite{zhang2015secure,zhang2021security,zhang2018game}.

Beyond these canonical forms, advanced game-theoretic models extend to environments with partial observability, bounded rationality, and adaptive learning. These extensions are essential in realistic cybersecurity settings, where agents operate with limited knowledge, may attempt to deceive one another, or continually refine their behavior based on observed outcomes.

\subsection{Game theoretic frameworks and Equilibrium Solutions}

Security games provide a rigorous formalism for analyzing strategic interactions among heterogeneous players operating in contested and uncertain environments. Within this framework, defenders, attackers, and benign users interact through well-defined action spaces and utility structures, subject to informational constraints and dynamic system evolution. A typical security game involves multiple strategic participants, most notably attackers, defenders, and compliant users. Let $N$ denote the number of players, with each player indexed by $i \in \mathcal{I} = \{1, 2, \ldots, N\}$. Each player $P_i$ is associated with a private type $\theta_i \in \Theta_i$, which encodes internal attributes such as threat level, behavioral bias, expertise, or operational role. These types introduce internal uncertainty into the system and critically shape how agents perceive their environment and formulate strategic choices.

The behavior of each player is governed by actions $a_i \in A_i$, chosen to achieve objectives aligned with their role, such as launching attacks, deploying defenses, or adhering to compliance protocols. Players may adopt either pure strategies, selecting a single action deterministically, or mixed strategies, wherein a probability distribution $\sigma_i \in \Delta A_i$ is placed over the action set. For defenders, typical actions include prevention mechanisms (e.g., segmentation, sandboxing), detection tools (e.g., intrusion detection systems, anomaly detectors), reactive responses (e.g., patching, containment), and proactive measures (e.g., deception, moving target defense, honeypots). Attackers, in contrast, typically follow a structured kill chain including reconnaissance, exploitation, privilege escalation, lateral movement, and data exfiltration. User behavior may reflect varying degrees of compliance, negligence, or insider threat risk.

Uncertainty in security games arises from both environmental and structural sources. External uncertainty stems from stochastic fluctuations in the system, such as unknown perturbations in state transitions or noisy observations. These are typically modeled as exogenous disturbances, $w_s$ and $w_o$, representing state and observation noise, respectively. Internal uncertainty originates from hidden player characteristics, specifically their types $\theta_i$, which are generally drawn from a common prior distribution $b_i^0$. This uncertainty is particularly pronounced in adversarial settings, where strategic deception and asymmetric information are prevalent.

Each player’s preferences are encoded in a utility function $u_i: A \times \Theta \times S \rightarrow \mathbb{R}$, which maps action profiles, types, and system states to real-valued payoffs. These utilities capture key performance metrics such as mission assurance, resource expenditure, operational effectiveness, and adversarial damage. Depending on the modeling context, utility functions may be single-objective (e.g., maximizing expected reward) or multi-objective (e.g., balancing risk, cost, and availability).

Information availability and rationality assumptions play a crucial role in shaping the strategic landscape. Players may have full, partial, or noisy observations of the game history and opponents' actions. The concept of common knowledge, the shared understanding of the structure and distributions of the game, may or may not hold. Players typically maintain Bayesian beliefs about other agents' types and update these beliefs in response to observed actions. Rationality assumptions vary across models, from perfectly rational players who best-respond to beliefs, to boundedly rational models that incorporate heuristics, satisficing, level-$k$ reasoning \cite{biswas2015measuring}, or prospect theory \cite{hu2023detection,kahneman2013prospect}.

The dynamic nature of the environment is modeled using a discrete-time framework. Let $s_k$ denote the system state at time $k$, and let $a_k = (a_1^k, \ldots, a_N^k)$ be the joint action taken by all players. The system evolves according to a transition function,
\[
s_{k+1} = f(s_k, a_k, \theta, w_s),
\]
which may incorporate endogenous factors (e.g., agent behavior) and exogenous randomness. Observations are defined by
\[
o_k^i = g_i(s_k, a_k, \theta, w_o),
\]
where $g_i$ captures the partial or noisy perception that player $i$ receives.

Each agent selects strategies that optimize their objectives over time, balancing short-term and long-term considerations. Objectives may be oriented toward immediate gains, such as one-shot payoffs (e.g., ransomware success), or long-horizon performance, such as discounted cumulative rewards. The spatial and functional scope of objectives also varies, ranging from localized defense of critical nodes to system-wide risk minimization. Players may adopt standard expected-utility criteria or instead seek robustness under uncertainty, employing risk-averse strategies that minimize worst-case losses.

\subsection{Classification of Security Games}

Security games can be systematically classified according to two fundamental axes: the temporal structure of interaction (static versus dynamic) and the completeness of information (complete versus incomplete). These dimensions yield distinct game-theoretic formulations and equilibrium concepts that govern strategic behavior.

\subsubsection*{Static Games with Complete Information}

In \emph{static security games with complete information}, the strategic interaction unfolds in a single stage, where all players choose their actions simultaneously without knowing the others' choices. The term 'static' emphasizes the absence of temporal evolution and 'complete information' indicates that all players are fully informed of the structure of the game, including the sets of players, the actions available for each player, and the utility functions that determine the rewards for all possible action profiles.

Formally, let $\mathcal{I} = \{1, 2, \ldots, N\}$ denote the finite set of players. Each player $i \in \mathcal{I}$ is associated with a finite action set $A_i$, and selects a \emph{mixed strategy} $\sigma_i \in \Delta(A_i)$, where $\Delta(A_i)$ denotes the space of probability distributions over $A_i$. A mixed strategy specifies the probability with which the player $i$ selects each action, allowing randomized behavior to prevent predictability, an essential feature in security settings.

The joint strategy profile is denoted $\sigma = (\sigma_1, \ldots, \sigma_N)$, and the joint action space is given by $A = A_1 \times \cdots \times A_N$. The utility of the player $i$ under a joint strategy profile $\sigma$ is defined as the expected value of the utility function $u_i: A \rightarrow \mathbb{R}$, computed as:
\begin{equation}
v_i(\sigma) = \sum_{a \in A} \left( \prod_{j=1}^N \sigma_j(a_j) \right) u_i(a),
\end{equation}
where $a = (a_1, \ldots, a_N)$ is a joint action profile, and the term $\prod_{j=1}^N \sigma_j(a_j)$ represents the probability that that profile occurs under mixed strategies.

A strategy profile $\sigma^* = (\sigma_1^*, \ldots, \sigma_N^*)$ is said to constitute a \emph{Nash Equilibrium} (NE) if no player has an incentive to unilaterally deviate, that is,
\begin{equation}
v_i(\sigma_i^*, \sigma_{-i}^*) \geq v_i(\sigma_i', \sigma_{-i}^*) \quad \forall \sigma_i' \in \Delta(A_i), \quad \forall i \in \mathcal{I},
\end{equation}
where $\sigma_{-i}^*$ denotes the strategies of all players other than $i$. This condition ensures that each player's strategy is a best response to the strategies of others.

In security contexts, static games with complete information often serve as a baseline model for analyzing scenarios where adversaries and defenders choose their actions once, such as selecting initial resource allocations or deciding on immediate threat responses. Although simplistic in temporal scope, these models reveal important strategic principles, including the potential for randomized defense policies (e.g., patrol randomization or randomized firewall rules) and the deterrent effects of credible threat modeling.

Nash equilibria in these games may be non-unique, and equilibrium selection often depends on additional refinements or assumptions, such as risk dominance, payoff dominance, or behavioral models. Furthermore, since all players know the utilities of others, these games are not suited to modeling deception or asymmetric information, which are instead captured by games with incomplete information.

\subsubsection*{Dynamic Games with Complete Information}

Dynamic security games generalize static formulations by incorporating temporal evolution and sequential decision-making. These games are particularly well-suited for modeling adversarial interactions that unfold over multiple stages, such as ongoing attacks, adaptive defenses, and evolving system conditions. In the complete information setting, all players are fully informed of the game structure, including the state transition dynamics, utility functions, and each other's action sets and available strategies.

The game is played over a finite or infinite discrete time horizon indexed by $k \in \{0, 1, 2, \ldots, K\}$, where $K$ may be finite or infinite. At each stage $k$, the system is in a state $s_k \in S$, drawn from a finite or continuous state space $S$. Players simultaneously observe the current state $s_k$ and then choose actions $a_k = (a_1^k, \ldots, a_N^k) \in A = A_1 \times \cdots \times A_N$, where $A_i$ denotes the action set of player $i$.

Unlike static games, dynamic security games incorporate \emph{stochastic system dynamics}, whereby the next state $s_{k+1}$ is determined probabilistically according to a known transition kernel:
\begin{equation}
\mathbb{P}(s_{k+1} \mid s_k, a_k) = f(s_{k+1}; s_k, a_k),
\end{equation}
where $f: S \times S \times A \rightarrow [0,1]$ defines the state transition probability from $s_k$ to $s_{k+1}$ given joint action $a_k$. This stochasticity models exogenous uncertainty arising from unpredictable system behavior, external events, or environmental noise.

Each player $i$ selects actions through a \emph{Markov policy} $\sigma_i^k: S \rightarrow \Delta A_i$, which maps the current state $s_k$ to a probability distribution over actions. The overall strategy profile is denoted $\sigma = (\sigma_1^0, \ldots, \sigma_N^K)$, with $\sigma_i^k \in \Delta(A_i)$ for each $i$ and $k$.

Given an initial state $s_{k_0}$ at time $k_0$, the cumulative expected utility for player $i$ over the horizon $[k_0, K]$ is defined as:
\begin{equation}
v_i(s_{k_0}, \sigma) = \mathbb{E}_{s_{k_0:K}, a_{k_0:K}} \left[ \sum_{k=k_0}^{K} \gamma^k u_i(s_k, a_k) \right],
\end{equation}
where $u_i: S \times A \rightarrow \mathbb{R}$ is the stage utility function, and $\gamma \in (0,1]$ is the discount factor that models temporal preferences, ensuring convergence in the infinite-horizon case.

The solution concept appropriate for this setting is the \emph{Subgame Perfect Nash Equilibrium} (SPNE). A strategy profile $\sigma^* = (\sigma_i^{k*})_{i \in \mathcal{I}, k \geq 0}$ constitutes an SPNE if, for every player $i \in \mathcal{I}$ and every time $k$, $\sigma_i^{k*}$ maximizes $v_i$ given the strategies of the other players and the current state:
\begin{equation}
\sigma_i^{k*}(s_k) \in \arg\max_{\sigma_i^k \in \Delta A_i} \, \mathbb{E} \left[ \sum_{t=k}^{K} \gamma^t u_i(s_t, a_t) \,\middle|\, s_k, \sigma_i^k, \sigma_{-i}^{*} \right].
\end{equation}

This equilibrium concept refines the notion of Nash equilibrium by requiring optimality at every subgame, i.e., at every state and time step, thereby enforcing \emph{intertemporal consistency}. It is particularly important in security settings, where adaptive adversaries and defenders must continuously revise strategies in response to unfolding events and changing system states.

Dynamic games with complete information are central to modeling a wide range of realistic security scenarios, including adversarial pursuit-evasion, defense resource reallocation, patrolling strategies, intrusion response planning, and cyber maneuver operations. The explicit modeling of stochastic dynamics allows for the inclusion of uncertainty due to system noise, random attacker behavior, and environmental volatility, key features of modern cyber-physical and sociotechnical systems \cite{zhu2025revisiting,zhu2020cross,tao22confluence}.

\subsubsection*{Dynamic Games with Incomplete Information}

Dynamic security games with incomplete information extend the multistage interaction framework to settings where players possess private information, commonly referred to as \emph{types}, that are not directly observable by others. These types encode internal characteristics such as intent, capability, risk tolerance, or strategic disposition. Let $\theta_i \in \Theta_i$ denote the private type of player $i \in \mathcal{I}$, and let $\theta = (\theta_1, \ldots, \theta_N) \in \Theta = \Theta_1 \times \cdots \times \Theta_N$ represent the full type profile.

At each time step $k$, the system state is denoted by $s_k \in S$, and it evolves stochastically according to a type-dependent transition kernel:
\begin{equation}\label{kernel}
\mathbb{P}(s_{k+1} \mid s_k, a_k, \theta) = f(s_{k+1}; s_k, a_k, \theta),
\end{equation}
where $a_k = (a_1^k, \ldots, a_N^k)$ is the joint action profile at stage $k$.

In this setting, players do not observe the types $\theta_{-i}$ of other participants. Instead, each player $i$ maintains a belief distribution $b_i^k \in \Delta(\Theta_{-i})$ over others' types, which is updated dynamically using Bayes' rule based on observable signals (e.g., actions, states, or messages). The player's \emph{information set} $H_k^i$ at time $k$ includes their own type $\theta_i$, the history of observable states and actions up to time $k$, and the belief $b_i^k$.

Thus, a behavioral strategy for the player $i$ at stage $k$ is defined as:
\begin{equation}
\sigma_i^k: H_k^i \rightarrow \Delta(A_i),
\end{equation}
mapping the current information history and private type to a distribution over actions.

The cumulative expected utility for player $i$ from time $k_0$ to $K$, given the strategy profile $\sigma = (\sigma_1^k, \ldots, \sigma_N^k)_{k=k_0}^{K}$ and initial state $s_{k_0}$, is expressed as:
\begin{equation}
v_i(s_{k_0}, \theta_i, \sigma) = \mathbb{E}_{\theta_{-i}, s_{k_0:K}, a_{k_0:K}} \left[ \sum_{k=k_0}^K \gamma^k u_i(s_k, a_k, \theta) \right],
\end{equation}
where $u_i: S \times A \times \Theta \rightarrow \mathbb{R}$ is the stage utility function, and $\gamma \in (0,1]$ is the discount factor.

A strategy profile $\sigma^* = (\sigma_i^{k*})_{i \in \mathcal{I}, k \geq 0}$ constitutes a \emph{Perfect Bayesian Nash Equilibrium} (PBNE) if the following conditions are satisfied:

\begin{enumerate}
    \item \textbf{Sequential Rationality:} For every player $i \in \mathcal{I}$, private type $\theta_i \in \Theta_i$, and information history $H_k^i$, the strategy $\sigma_i^{k*}(H_k^i)$ maximizes the expected utility $v_i$ given the current belief $b_i^k$ and the strategies $\sigma_{-i}^{*}$ of the other players.
    \item \textbf{Belief Consistency:} The belief distribution $b_i^k$ is updated according to Bayes' rule wherever applicable, using the observed sequence of public information and assuming that all players adhere to the equilibrium strategy profile $\sigma^*$.
\end{enumerate}

PBNE serves as the canonical solution concept for dynamic games under asymmetric information. It tightly integrates belief updating with strategic decision-making, allowing each player to reason about others' private information through observed behavior. This structure is particularly powerful for modeling deception, trust dynamics, learning in adversarial environments, and strategic signaling in cybersecurity and multi-agent security operations.

Such games are foundational in the study of advanced security problems including multi-stage attacks, insider threat detection, strategic misinformation, and adaptive adversarial inference, where uncertainty about other agents’ objectives or capabilities is a defining challenge.

\subsection{Emerging Solution Concepts}

The classical solution concept in game theory is the equilibrium, which serves as a foundational tool for analyzing strategic behavior. Equilibrium concepts—such as Nash equilibrium or Stackelberg equilibrium—are particularly valuable in cybersecurity for risk assessment, designing defense and attack strategies, and mechanism design that aligns agent incentives with desired system outcomes. By identifying stable strategy profiles where no agent has an incentive to unilaterally deviate, equilibrium analysis offers predictive power and supports principled decision-making under adversarial conditions.

However, equilibrium-based analysis may be insufficient in many cybersecurity settings, particularly when interactions are nonstationary, bounded in time, or subject to information asymmetry and adaptation. In dynamic environments where threats evolve, systems are reconfigured, and agents continuously learn, the assumption of convergence to equilibrium may not hold. This is especially relevant in real-world cyber engagements, where interactions are episodic or transient and agents often act under uncertainty with limited information or foresight.

In such cases, alternative solution concepts, such as dynamic games with finite horizon, learning dynamics, or regret minimizing strategies, may be more appropriate \cite{tao22confluence}. These frameworks emphasize adaptive behavior, temporal evolution, and robust responses to unpredictable or rapidly changing adversaries. Incorporating these perspectives into cybersecurity models allows for a more realistic and flexible understanding of adversarial dynamics, enabling defense mechanisms that are resilient not just to known threats but also to emerging and evolving ones.

\subsection{Applications of Security Games}

Game theory offers a principled way to model adversarial interactions, anticipate attacker behavior, and design effective defense mechanisms. Several key paradigms within cybersecurity illustrate the critical role of game-theoretic methods:

\subsubsection*{Game-Theoretic Modeling of Cyber Deception}

Cyber deception exploits the asymmetry of information to mislead adversaries through deliberate manipulation of signals, observations, and environmental responses. Game-theoretic frameworks, particularly \emph{signaling games} and \emph{Stackelberg games}, provide a principled foundation for modeling such interactions in adversarial environments.

\paragraph{Signaling Game Framework}

A one-shot signaling game involves a defender (sender) and an attacker (receiver). Nature draws the true system type \( \theta \in \Theta = \{\theta_1, \theta_2\} \), such as a real system (\( \theta_1 \)) or a honeypot (\( \theta_2 \)), from a known prior \( P(\theta) \). The defender, who observes \( \theta \), selects a signal \( s \in S \) according to a strategy \( \sigma_D: \Theta \rightarrow \Delta(S) \). The attacker observes \( s \), updates their belief \( \mu(\theta \mid s) \) using Bayes’ rule, and selects an action \( a \in A \) via a strategy \( \sigma_A: S \rightarrow \Delta(A) \). 

The utilities \( u_D(\theta, s, a) \) and \( u_A(\theta, s, a) \) represent the payoffs of the defender and attacker, respectively. A strategy profile \( (\sigma_D^*, \sigma_A^*) \) forms a \emph{Perfect Bayesian Nash Equilibrium (PBNE)} if:
\begin{enumerate}
    \item \textbf{Belief Consistency:} The belief \( \mu(\theta \mid s) \) is updated via Bayes’ rule wherever possible;
    \item \textbf{Sequential Rationality:} Given beliefs, each player optimizes their expected utility:
    \begin{align}
        \sigma_A^*(s) &\in \arg\max_{a \in A} \sum_{\theta \in \Theta} \mu(\theta \mid s) u_A(\theta, s, a), \\
        \sigma_D^*(\theta) &\in \arg\max_{s \in S} u_D(\theta, s, \sigma_A^*(s)).
    \end{align}
\end{enumerate}

\paragraph{Dynamic Stackelberg Game for Deception}

In repeated or multi-stage settings, deception is more effectively modeled using a \emph{dynamic Stackelberg game} with incomplete information. Let the system evolve over discrete time steps \( k = 0, 1, \ldots \), with state \( s_k \in S \) and joint action profile \( a_k = (a_k^D, a_k^A) \in A_D \times A_A \), where \( a_k^D \) and \( a_k^A \) are the actions of the defender and attacker, respectively.

The state transition is governed by a stochastic kernel as in (\ref{kernel}), 
and the defender and attacker optimize their respective cumulative utilities:
\begin{align}
J_D &= \mathbb{E} \left[ \sum_{k=0}^\infty \gamma^k u_D(s_k, a_k, \theta) \right], \\
J_A &= \mathbb{E} \left[ \sum_{k=0}^\infty \gamma^k u_A(s_k, a_k, \theta) \right],
\end{align}
where \( \gamma \in (0, 1] \) is a discount factor and \( \theta = (\theta_D, \theta_A) \in \Theta \) represents the type profile.

The defender (leader) commits to a policy \( \sigma_D: H_k^D \rightarrow \Delta(A_D) \), where \( H_k^D \) denotes the defender's observation history. The attacker (follower) observes the defender’s actions or signals and selects a best response strategy \( \sigma_A^*(\sigma_D) \). The defender’s problem becomes:
\begin{equation}
\sigma_D^* \in \arg\max_{\sigma_D} \mathbb{E} \left[ \sum_{k=0}^\infty \gamma^k u_D(s_k, \sigma_D(H_k^D), \sigma_A^*(H_k^A)) \right],
\end{equation}
where \( H_k^A \) is the attacker’s information set, which may include signals, actions, or partial state observations.

Cyber deception games form a foundational class of security games in which defenders exploit informational asymmetry to manipulate attacker perception and behavior \cite{pawlick2019game}. Game-theoretic models, particularly signaling games \cite{pawlick2018modeling}, Stackelberg games \cite{yang2025deceive,pawlick2021game}, and partially observable stochastic games (POSGs) \cite{horak2017manipulating}, enable principled design and analysis of deception mechanisms. In these models, the defender strategically crafts signals or system responses to induce belief shifts in the attacker’s inference process, denoted \( b_k^A \), ultimately steering adversarial actions toward less damaging outcomes.

Practical implementations of such strategies include honeypots \cite{spitzner2002honeypots,qian2020receding}, decoy systems \cite{kulkarni2020decoy,nasr2016game}, moving target defenses (MTD) \cite{jajodia2011moving,zhu2013game}, and obfuscation techniques \cite{fang2020channel,pawlick2016stackelberg,fang2021fundamental}. These deception tools aim to:
\begin{itemize}
    \item Lure attackers into interacting with monitored or inert components;
    \item Induce attackers to expend resources on misrepresented vulnerabilities;
    \item Delay or mislead adversarial planning, reducing operational risk.
\end{itemize}

\emph{Detection games} constitute a key subclass of deception games \cite{hu2022evasion,hu2024game}. Here, the defender aims to detect malicious activity, while the attacker seeks to evade detection. The defender chooses detection resource allocations, such as sensor placements or intrusion detection policies, while the attacker selects stealthy attack vectors. The interaction is typically modeled as a partially observable or Bayesian game, where the attacker’s presence or type is hidden, and deception arises through the defender’s attempt to infer and expose this hidden state.

Similarly, \emph{jamming games} represent another important subclass of deception games, particularly in wireless and cyber-physical systems \cite{zhu2010stochastic,zhu2011eavesdropping,xu2017game,nugraha2019subgame}. In these settings, a jammer (attacker) seeks to degrade communication performance by introducing interference, while the defender employs adaptive transmission strategies, such as frequency hopping, power control, or route diversity, to maintain system reliability. Jamming deception arises as the defender masks transmission patterns or dynamically shifts strategies to confuse or neutralize the jammer. These interactions are often captured as stochastic zero-sum or general-sum games with energy constraints and noisy feedback loops.

\subsubsection*{Game-Theoretic Modeling of Network Security Games}

Network security games constitute an important class of security games in which the underlying structure of the system is represented as a graph \( G = (V, E) \), where nodes \( V \) and links \( E \) correspond to critical assets such as hosts, routers, communication paths, or physical infrastructure components. In these games, the defenders aim to protect, reinforce, or monitor subsets of the network, while the attackers seek to compromise, disable, or remove key elements to degrade the overall functionality of the system.

A prominent subclass is the \emph{network interdiction game} \cite{8673619,chen2021dynamic,chen2019control,nugraha2020dynamic,nugraha2025rolling}, which models the strategic interaction between an interdictor (attacker) and a defender over the network topology. The attacker selects a set of nodes or edges to remove or disable, subject to resource or budget constraints. The defender, in turn, allocates protection resources to harden certain components or ensure survivability through redundancy. The primary objective of the attacker is often to maximize disruption, e.g., by disconnecting critical paths, increasing latency, or degrading flow throughput, while the defender aims to minimize such impact.

Formally, let \( x \in \{0,1\}^{|E|} \) denote the attacker's interdiction strategy, where \( x_e = 1 \) indicates that edge \( e \in E \) is targeted for removal, and let \( y \in \{0,1\}^{|E|} \) represent the defender’s allocation of protection resources. The network performance metric, such as maximum flow, shortest path length, or reachability, denoted \( \phi(G; x, y) \), is a function of both strategies. The defender solves:
\begin{equation}
\min_{y \in \mathcal{Y}} \max_{x \in \mathcal{X}} \, \phi(G; x, y),
\end{equation}
where \( \mathcal{X} \) and \( \mathcal{Y} \) are feasible action sets under budget or cardinality constraints.

Such games are inherently hierarchical and are often formulated as \emph{bilevel optimization problems}, reflecting the sequential nature of decision-making. The attacker (leader) may act first in a Stackelberg framework, or both players may move simultaneously in a zero-sum game setting. Extensions also account for uncertainty in attacker capabilities, stochastic network states, and partial observability of attacks.

Network interdiction games are widely applied in domains including military logistics \cite{farina2025contested}, cyber-physical infrastructure protection \cite{chen2019game,chen2019game}, and supply chain security \cite{nagurney2015supply,kieras2022iot}. They support strategic planning for cyber resilience by identifying critical nodes and links, evaluating system vulnerabilities, and optimizing the deployment of limited defensive resources under adversarial threat models.

\subsubsection*{Game-Theoretic Modeling of Socio-Economic Security Games}

Beyond purely technical domains, human actors play a central role in shaping cybersecurity outcomes. Game-theoretic models are particularly well suited to capture the complex interactions arising from cognitive limitations, behavioral biases, trust dynamics, and organizational incentives. These models extend the traditional framework of security games to account for socio-economic and psychological dimensions of security decision-making.

Human participants, including users, system administrators, defenders, and even adversaries, introduce uncertainties that stem not only from incomplete information but also from bounded rationality and subjective preferences. In contrast to fully rational agents, boundedly rational actors may rely on heuristics, exhibit framing effects, or display risk-sensitive behavior that varies across context \cite{rass2020bounded,chen2018security,yang2023game}. These behavioral attributes are particularly salient in phishing, social engineering, and insider threat scenarios, where attackers exploit cognitive vulnerabilities and patterned decision-making.

Game-theoretic approaches to human-centered security often incorporate non-classical belief models, such as:
\begin{itemize}
    \item \emph{Level-$k$ reasoning}, where agents reason about others at varying levels of strategic depth;
    \item \emph{Quantal response equilibria}, modeling probabilistic action selection based on perceived payoffs;
    \item \emph{Prospect theory}-based utilities, capturing non-linear valuation of gains and losses.
\end{itemize}

In organizational settings, multilevel game models can represent the interactions among policymakers, technical staff, and end-users, each with distinct utility functions and information sets. These models enable analysis of how security policies, incentives, or training interventions affect compliance and behavior. For example, behavioral security games can evaluate the effectiveness of nudges, penalties, or educational campaigns in shifting user choices toward more secure equilibria.

Furthermore, adversarial modeling of trust and perception is increasingly critical in the presence of AI-generated content, misinformation campaigns, and automated social manipulation. Repeated games with incomplete information provide a natural framework for modeling trust evolution, where players form and update beliefs based on historical actions and observed signals. Attackers may strategically manipulate these beliefs to erode the integrity or credibility of the system over time, requiring defenses that are both technically robust and psychologically resilient.

\section{Moving Toward LLM Agents}

LLMs represent a major step forward in artificial intelligence, offering powerful capabilities to understand, generate, and reason over natural language. Unlike traditional rule-based or narrowly trained machine learning systems, LLMs are trained on vast corpora of text, allowing them to encode patterns of human communication, conceptual reasoning, and strategic thinking. They can generate coherent text, perform complex reasoning tasks, simulate perspectives, and interact with humans and other agents in dynamic environments. These capabilities make LLMs not only valuable tools for automation, but also foundational building blocks for constructing agentic AI systems: autonomous agents capable of perception, reasoning, communication, and action.

\subsection{LLMs as a Catalyst for Advancing Game-Theoretic Cybersecurity Models}
LLMs offer an opportunity to overcome several foundational limitations in classical game-theoretic frameworks, particularly in cybersecurity contexts, where strategic interactions are dynamic, uncertain, and often human-in-the-loop. Below, we elaborate on key bottlenecks in classical game theory and how LLMs offer new conceptual and operational pathways forward.

\paragraph{Rational Decision-Making Assumptions.}
Classical game theory models agent behavior as the outcome of a utility-maximizing optimization process. This optimization-based reasoning is grounded in foundational theories such as von Neumann and Morgenstern's expected utility theory and Savage's axioms. However, such models assume fully rational agents with complete knowledge of their preferences and action spaces. The reasoning process is externally imposed and internalized through maximization.

LLMs, by contrast, do not rely on predefined utility functions. Instead, they produce decisions and responses based on contextual prompts, language cues, analogical reasoning, and prior exemplars, offering a more flexible and descriptive model of reasoning. Classical optimization can be interpreted as a phenomenological abstraction of rational behavior, while LLMs enable a cognitively plausible account of how reasoning unfolds in settings characterized by ambiguity, competing goals, or adaptive constraints.

\paragraph{Knowledge Assumptions and Epistemic Limitations.}
Traditional game-theoretic models often rely on the assumption of common knowledge: all players know the structure of the game, know that others know it, and so on. Although analytically convenient, this assumption does not capture \emph{epistemic heterogeneity} differences in agents' beliefs, knowledge, or awareness of one another’s capabilities. In cybersecurity, where information asymmetry is pervasive and perspectives are misaligned, such heterogeneity is especially critical \cite{pawlick2021game}.

LLMs provide a mechanism for simulating epistemic diversity by generating beliefs, decisions, and justifications based on partial, role-specific, or even inconsistent information. This enables a novel form of \emph{epistemic simulation}, where agents reason from divergent worldviews, enriching the modeling of strategic uncertainty.

\paragraph{Limitations of Classical Solution Concepts.}
Game-theoretic solution concepts such as the Nash equilibrium or the Stackelberg equilibrium are foundational, but often rely on stringent assumptions: stationary environments, complete knowledge, and rational foresight. In cybersecurity, interactions are often non-stationary, episodic, and adaptive. Agents must act under time constraints and informational incompleteness, and equilibrium solutions may not exist or be computationally tractable.

LLMs suggest new forms of equilibrium grounded in reasoning traces, interactive language, and adaptive simulation. Instead of fixed-point solutions, LLMs can support \emph{LLM-based equilibria}, in which agent strategies co-evolve through dialogue, recursive prompting, or context-sensitive generation. These equilibria reflect dynamic adaptability rather than static optimality.

\paragraph{Language-Based Signaling and Interpretation.}
Classical signaling games model communication through discrete symbols or numerical messages. However, strategic communication in cybersecurity, especially in phishing, negotiation, or deception, is based on natural language, which carries ambiguity, context dependence, and pragmatic nuance.

LLMs allow for the generation, interpretation, and adaptation of natural language messages within game-theoretic settings. This makes it possible to model linguistic deception, persuasion, and framing effects in a native medium, rather than abstracting them into symbolic proxies. Thus, LLMs support a richer modeling of \emph{semiotic interactions}, where the meaning of a message is inferred from content, form, and context.

\paragraph{Modeling Human Bounded Rationality.}
Classical models frequently assume that humans behave as perfectly rational utility maximizers. In reality, human decisions are shaped by bounded rationality, cognitive biases, emotional states, and contextual heuristics. Capturing these behaviors within game-theoretic models is challenging and often requires oversimplifying assumptions.

LLMs, trained on large-scale human-generated corpora, can simulate human-like reasoning and behavior. They reflect linguistic, cognitive, and social patterns, enabling realistic modeling of users, insiders, and attackers. This makes LLMs powerful tools for designing human-centered defenses, performing red teaming, or evaluating susceptibility to manipulation or misinformation.

\subsection{Contextual Reasoning and Epistemic Uncertainty in Agentic AI for Cybersecurity}

In addition to the foundational capabilities discussed previously, LLMs offer advanced capacities for \emph{contextual reasoning} and handling \emph{epistemic uncertainty}, both of which are critical in adversarial and dynamically evolving cyber environments.

\paragraph{Use of Contextual Information and External Tools.}
One of the defining capabilities of agentic AI systems is their ability to augment internal reasoning with dynamic access to external contextual information and operational tools. In contrast to traditional machine learning models, which are trained on fixed datasets and operate primarily through static inference, AI architectures support real-time interaction with their environment. This includes querying APIs, invoking automated analysis services, and integrating heterogeneous data sources as part of the decision-making process.

For instance, consider a scenario where a DNS query is issued to a suspicious domain, such as \texttt{abcxyz.ru}. A conventional machine learning classifier may lack the contextual knowledge necessary to evaluate the threat if the domain is not present in its training data. In contrast, an agentic AI system can execute a multi-step reasoning process: it may query VirusTotal, consult AbuseIPDB, cross-reference domain reputation across threat intelligence feeds, and correlate findings with internal network telemetry. Based on the aggregate confidence of these contextual signals, the system can autonomously decide to quarantine the requesting host, escalate to a human analyst, or continue passive observation while updating its internal threat model.

This paradigm mirrors the behavior of a seasoned cybersecurity analyst, capable of synthesizing distributed knowledge, applying domain-specific heuristics, and adapting responses based on evolving context. However, the agentic system achieves this with the precision and speed of automation and at a scale that exceeds human capacity. Such an integration of external tools and intelligence sources allows agentic AI to operate with situational awareness and adaptive responsiveness, critical traits for effective cyber defense in complex environments.

\paragraph{Prediction and Management of Epistemic Uncertainty.}
A particularly compelling advantage of agentic AI lies in its ability to reason under epistemic uncertainty, uncertainty arising from incomplete or unmodeled knowledge, particularly in the presence of zero-day exploits and advanced persistent threats (APTs) \cite{zhu2018multi}. These threats are often engineered to circumvent signature-based detection by leveraging novel techniques or exploiting legitimate system functions (commonly known as "living-off-the-land" attacks). Classical detection models, which require labeled training data or pre-defined rules, are inherently limited in their ability to identify such previously unseen behaviors.

Agentic AI addresses this limitation by incorporating abductive and counterfactual reasoning. Rather than relying solely on pattern matching, the system can generate explanatory hypotheses for anomalous observations, simulate potential threat scenarios, and evaluate their plausibility based on background knowledge, current telemetry, and historical behavior. For example, suppose that the agent observes an unusual pattern of system calls on an industrial control system, such as a rarely invoked kernel module interacting with a programmable logic controller (PLC). While this sequence may not match any known attack signature, the agent can hypothesize a range of possibilities, including legitimate maintenance operations or lateral movement by a stealthy adversary.

By evaluating these hypotheses in parallel, agentic AI can compute risk-weighted projections of future states and proactively trigger mitigations. This could include temporarily isolating the affected subnet, stopping critical operations, or launching further diagnostic probes, all before definitive confirmation is available. In doing so, agentic AI operationalizes a form of anticipatory defense with risk awareness that is fundamentally distinct from traditional reactive approaches.

\paragraph{Toward Cognitively Informed Cyber Defense.}
The capabilities described above highlight the shift from pattern-based classification to cognition-inspired strategic reasoning in cyber defense. Agentic AI systems integrate symbolic and sub-symbolic reasoning, language understanding, external tool invocation, and multi-hypothesis evaluation, enabling them to function as autonomous cyber analysts. Their ability to dynamically contextualize events, manage epistemic uncertainty, and infer causality makes them particularly well suited for adversarial settings marked by deception, incomplete information, and real-time constraints.

\section{Modeling LLM Agents}
LLMs fundamentally transform how we model decision-making in multi-agent systems, particularly in strategic contexts such as cybersecurity. Traditional game-theoretic formulations often rely on the assumption that agents make decisions by solving a utility maximization problem, formally represented as $\arg\max_{a} u(a, I)$, where the agent selects the action $a$ that yields the highest utility given the information $I$. This optimization-driven reasoning process is well justified in classical decision theory, but it imposes strong assumptions about rationality, completeness of knowledge, and cognitive uniformity that rarely hold in practice.

LLMs offer a new paradigm that relaxes these assumptions by enabling agents to generate decisions not as the output of an explicit optimization problem, but as responses conditioned on observed context and latent internal representations. In this view, an action distribution $\mu(a)$ is produced by an LLM conditioned on the agent’s information $I$ and an internal, flexible state $\hat{x}$. This representation $\hat{x}$ is not rigidly defined: it is rich, high-dimensional, and capable of capturing more than just goal-oriented behavior. It includes elements of intent, perception, and inference about others, allowing for forms of decision-making that are more expressive and adaptive.

This flexibility enables LLMs to articulate not only the \textit{goal} and the \textit{capability} of an agent but also the \textit{reasoning process} itself. Whereas classical models hardcode the reasoning through an optimization structure, LLMs encode patterns of reasoning phenomenologically, by simulating how reasoning has been observed and recorded in language. This results in richer and more varied forms of agent behavior, better suited to environments where assumptions about perfect rationality or complete knowledge break down.

Another critical advancement that LLMs bring is their ability to handle epistemic asymmetries. Classical game theory often assumes \textit{common knowledge} of the structure of the game, including the knowledge and beliefs of all players ad infinitum. In reality, especially in adversarial settings like cybersecurity, agents frequently operate under \textit{asymmetric information}: different agents possess different knowledge, perspectives, or mental models. LLMs are naturally suited to simulate such asymmetries, since they can condition outputs on partial, conflicting, or role-specific information and reason in ways that are non-uniform across agents.

Furthermore, LLMs enable recursive and social reasoning, including \textit{reasoning about other agents’ reasoning}. This capacity can be expressed through models where an agent generates action distributions $\mu(a \mid I_j, x)$ based on an understanding of the knowledge and context of other agents. This supports more realistic modeling of inter-agent interactions, especially when such interactions involve signaling, deception, or theory-of-mind dynamics.

In addition to these cognitive and epistemic advantages, LLMs also allow for \textit{linguistically grounded modeling}. In traditional signaling games, messages are often abstracted as discrete symbols or real-valued numbers. LLMs, however, can generate and interpret full natural language messages, making it possible to model signaling and persuasion in a way that mirrors actual human communication. This opens up new avenues for modeling language-based interaction strategies, such as adversarial prompts, misinformation, and adaptive deception.

Lastly, LLMs provide a natural interface for modeling \textit{bounded rationality}, the idea that human agents do not optimize perfectly but instead follow heuristics, are influenced by framing effects and often behave inconsistently. Drawing on patterns of human behavior encoded in the training data, LLMs can simulate these imperfections more faithfully than normative models. As such, they are closer to modeling how humans actually reason, especially under stress, uncertainty, or cognitive load.

\subsection{LLM-Based Nash Game: Prompt-Driven Reasoning and Strategic Equilibrium}

The LLM-based Nash game generalizes classical game-theoretic formulations by shifting the locus of strategic decision-making from action-level utility maximization to \emph{reasoning-level prompt selection}. In this framework, agents employ LLMs as generative policy mechanisms, mapping structured prompts and local information into probabilistic strategies. This is especially salient in domains such as cybersecurity, where agents must reason under uncertainty, adapt to incomplete information, and engage in strategic behavior shaped by perception and belief.

Consider a simultaneous-move game between two agents: an attacker \( A \) and a defender \( D \). The environment is defined by a standard normal-form structure:
\(
\mathcal{G} = (\mathcal{A}, \mathcal{D}, u_A, u_D),
\)
where \( \mathcal{A} \) and \( \mathcal{D} \) denote the action spaces of the attacker and defender, and \( u_A, u_D : \mathcal{A} \times \mathcal{D} \to \mathbb{R} \) are their respective utility functions. Each agent is endowed with a \emph{mindset} \( M_A = (I_A, \mathcal{X}, \theta) \) and \( M_D = (I_D, \mathcal{Y}, \delta) \), where \( I_A \), \( I_D \) are private information sets, \( \mathcal{X}, \mathcal{Y} \) are structured prompt spaces, and \( \theta, \delta \) denote internal model parameters such as worldviews induced by the LLMs.

Rather than choosing actions directly, each agent queries its LLM with a prompt \( x \in \mathcal{X} \) or \( y \in \mathcal{Y} \), conditioned on private information, to generate a stochastic policy:
\[
\mu_A(a \mid I_A, x) = \gamma_A(a \mid I_A, x, \theta), \quad 
\mu_D(d \mid I_D, y) = \gamma_D(d \mid I_D, y, \delta),
\]
where \( \gamma_A \) and \( \gamma_D \) are the generative reasoning-to-action mappings defined by the respective LLMs. These mappings reflect the agent's bounded rationality and cognitive constraints.

 In contrast to classical Nash equilibria, which operate on action spaces, the \emph{LLM-Nash equilibrium} is defined over the reasoning prompt space. The objective of each agent is to select a prompt, i.e., a cognitive strategy, that induces a favorable action distribution through its LLM.

\begin{definition}[LLM-Nash Reasoning Equilibrium]
Let \( (x^*, y^*) \in \mathcal{X} \times \mathcal{Y} \) be a pair of reasoning prompts. Then \( (x^*, y^*) \) is an LLM-Nash equilibrium if:
\begin{align}
\mathbb{E}_{a \sim \mu_A(\cdot \mid I_A, x^*),\, d \sim \mu_D(\cdot \mid I_D, y^*)}[u_A(a, d)] 
&\geq \mathbb{E}_{a \sim \mu_A(\cdot \mid I_A, x'),\, d \sim \mu_D(\cdot \mid I_D, y^*)}[u_A(a, d)] \quad \forall x' \in \mathcal{X}, \\
\mathbb{E}_{a \sim \mu_A(\cdot \mid I_A, x^*),\, d \sim \mu_D(\cdot \mid I_D, y^*)}[u_D(a, d)] 
&\geq \mathbb{E}_{a \sim \mu_A(\cdot \mid I_A, x^*),\, d \sim \mu_D(\cdot \mid I_D, y')}[u_D(a, d)] \quad \forall y' \in \mathcal{Y}.
\end{align}
\end{definition}

The policies \( (\mu_A^*, \mu_D^*) \) induced by \( (x^*, y^*) \) are referred to as the \emph{LLM-Nash behavioral equilibrium}. Crucially, this formulation places strategic reasoning, not just final decisions, at the center of equilibrium analysis.

Each prompt represents a deliberate act of reasoning, encoding the agent’s mental model of the game, beliefs about the opponent, and heuristics for decision-making. Thus, prompt selection functions as a metastrategy, aligning with the agent's epistemic constraints and representational capabilities. The equilibrium concept captures this alignment by ensuring no agent has an incentive to deviate in its reasoning design.

 \subsubsection*{Example: Reasoning Equilibria in Extended Prompt-Space RPS}

Consider a zero-sum Rock-Paper-Scissors (RPS) game between two agents, the attacker \( A \) and the defender \( D \), each with identical action spaces:
\(
\mathcal{A} = \mathcal{D} = \{\texttt{Rock}, \texttt{Paper}, \texttt{Scissors}\}.
\)
The payoff matrix for player \( A \) is defined as
\[
U_A = 
\begin{bmatrix}
0 & -1 & 1 \\
1 & 0 & -1 \\
-1 & 1 & 0
\end{bmatrix},
\]
with the defender's utility given by \( u_D = -u_A \) due to the zero-sum structure.

Each agent selects a prompt from their respective reasoning space. The attacker chooses from \( \mathcal{X} = \{x_1, x_2, x_3, x_4, x_5\} \), while the defender selects from \( \mathcal{Y} = \{y_1, y_2, y_3, y_4, y_5\} \). These prompts are interpreted as follows: \( x_1 \) instructs the attacker to exploit the opponent’s bias; \( x_2 \) assumes uniform play; \( x_3 \) counters the opponent’s last move; \( x_4 \) minimizes risk based on historical variance; and \( x_5 \) mimics the opponent’s observed behavior. On the defender’s side, \( y_1 \) randomizes equally; \( y_2 \) attempts to exploit the attacker’s last move; \( y_3 \) responds proportionally to the attacker’s empirical frequencies; \( y_4 \) selects the safest option based on past loss data; and \( y_5 \) anticipates potential bait and counters it.

The LLMs generate the following action distributions in response to the prompts. For the attacker: 
\[
\mu_A(x_1) = (0.2, 0.6, 0.2), \quad
\mu_A(x_2) = \left(\tfrac{1}{3}, \tfrac{1}{3}, \tfrac{1}{3}\right), \quad
\mu_A(x_3) = (0.4, 0.3, 0.3),
\]
\[
\mu_A(x_4) = (0.25, 0.5, 0.25), \quad
\mu_A(x_5) = (0.3, 0.4, 0.3).
\]
For the defender:
\[
\mu_D(y_1) = \left(\tfrac{1}{3}, \tfrac{1}{3}, \tfrac{1}{3}\right), \quad
\mu_D(y_2) = (0.3, 0.4, 0.3), \quad
\mu_D(y_3) = (0.2, 0.6, 0.2),
\]
\[
\mu_D(y_4) = (0.25, 0.5, 0.25), \quad
\mu_D(y_5) = (0.3, 0.2, 0.5).
\]

The expected utility for the attacker, denoted \( U_{ij} = \mathbb{E}[u_A(x_i, y_j)] \), is computed as the bilinear form \( \mu_A^\top U_A \mu_D \). For example, consider the pair \( (x_1, y_3) \). The attacker uses prompt \( x_1 \), generating \( \mu_A = (0.2, 0.6, 0.2) \), while the defender responds with \( \mu_D = (0.2, 0.6, 0.2) \) under prompt \( y_3 \). The expected utility is computed as
\[
U_{13} = (0.2, 0.6, 0.2)
\begin{bmatrix}
0 & -1 & 1 \\
1 & 0 & -1 \\
-1 & 1 & 0
\end{bmatrix}
\begin{bmatrix}
0.2 \\ 0.6 \\ 0.2
\end{bmatrix}
= 0.
\]

As a second example, for the prompt pair \( (x_3, y_5) \), we have
\[
\mu_A = (0.4, 0.3, 0.3), \quad \mu_D = (0.3, 0.2, 0.5),
\]
leading to
\[
U_{35} = \mu_A^\top U_A \mu_D = 0.04.
\]

For a third case, both players choose conservative prompts: \( x_4 \) and \( y_4 \), respectively. Each generates the same distribution \( (0.25, 0.5, 0.25) \), and the expected utility is again zero:
\[
U_{44} = (0.25, 0.5, 0.25)^\top U_A (0.25, 0.5, 0.25) = 0.
\]

Assume empirical evaluations indicate that the attacker achieves the highest average expected utility when using prompt \( x_5 \), and that the defender’s best response is prompt \( y_3 \). Then, with 
\[
\mu_A(x_5) = (0.3, 0.4, 0.3), \quad \mu_D(y_3) = (0.2, 0.6, 0.2),
\]
we obtain
\[
U_{53} = \mu_A^\top U_A \mu_D = 0.02.
\]

If neither player can improve their expected utility by unilaterally changing prompts, the pair \( (x_5, y_3) \) constitutes a reasoning-level equilibrium. The resulting behavioral equilibrium is
\[
\mu_A^* = (0.3, 0.4, 0.3), \quad \mu_D^* = (0.2, 0.6, 0.2),
\]
which deviates from the classical Nash equilibrium of uniform randomization. This deviation illustrates that when agents are constrained by fixed LLM architectures and bounded prompt spaces, they may settle into stable cognitive equilibria that are rational within their reasoning models, though potentially suboptimal in the classical game-theoretic sense.

The LLM-Nash game allows for the design and control of multi-agent systems by intervening in the agents’ \emph{reasoning space}, e.g., modifying prompts, providing auxiliary information, or expanding the cognitive scaffolding (mindsets). This enables a novel form of mechanism design grounded in \emph{prompt space engineering}, where desirable behaviors emerge from shaping the reasoning process, not merely manipulating payoffs.

\subsection{LLM-based Stackelberg Game}

Classical signaling games provide a foundational framework for modeling strategic communication under asymmetric information. In the canonical form, a sender observes private information \( I_S \) and selects a message \( m \in \mathcal{M} \) to transmit to a receiver, who, upon observing \( m \) and incorporating their own private information \( I_R \), selects an action \( a \in \mathcal{A} \). The outcome of the interaction is evaluated through a joint utility function \( u_S(m, a; I_S, I_R) \) and \( u_R(m, a; I_S, I_R) \), reflecting the incentives of the sender and receiver respectively. Equilibrium concepts such as Bayesian or Stackelberg equilibria typically rely on optimization over well-defined, symbolic strategy spaces.

However, this classical abstraction, while elegant, imposes severe limitations on the expressivity of communication and reasoning. Finite message spaces, static utility assumptions, and fixed-form best-response mappings cannot capture the richness of natural language, the cognitive dynamics of belief formation, or adaptive reasoning under uncertainty. To address these constraints, we introduce \emph{LLM-Stackelberg game}, where both the sender and the receiver are agents enabled by LLM engaging in prompt-driven generative reasoning.

In this augmented framework, the sender (acting as the Stackelberg leader) constructs a reasoning prompt \( x \in \mathcal{X} \) based on its private observation \( I_S \). This prompt, which may include chain-of-thought logic, intent encoding, or historical adaptation, is used to condition the sender’s LLM to produce a distribution over messages:
\[
\mu_S(m \mid I_S, x) = \mathbb{P}_{\texttt{LLM}_S}[m \mid I_S, x].
\]
The receiver, upon receiving the message \( m \), constructs its own reasoning prompt \(  y(m, I_R) \in \mathcal{Y} \), which incorporates both the observed signal and the receiver’s local context. The LLM-based action policy is then given by:
\[
\mu_R(a \mid m, I_R, y) = \mathbb{P}_{\texttt{LLM}_R}[a \mid m, I_R, y].
\]

This formulation introduces a two-layer decision structure: at the \textit{behavioral level}, the LLMs generate policies \( \mu_S \) and \( \mu_R \) based on prompts \( x \) and \( y \); at the \textit{reasoning level}, the agents optimize over their respective prompt spaces to influence downstream generative outcomes. The Stackelberg nature of the game manifests itself in the anticipatory optimization of the sender: the sender selects a prompt \( x \in \mathcal{X} \) with the foresight of the receiver's best response, which itself depends on \( m \), \( I_R \), and \( y \). The expected utility of the sender is thus expressed as:
\[
\bar{U}_S(x) := \mathbb{E}_{m \sim \mu_S(\cdot \mid I_S, x)}\left[
\max_{y \in \mathcal{Y}} \; \mathbb{E}_{a \sim \mu_R(\cdot \mid m, I_R, y)} \left[ u_S(m, a) \right]
\right],
\]
where the inner maximization over \( y \) reflects the receiver’s own reasoning optimization.

The receiver, for a fixed message \( m \), chooses a prompt \( y \in \mathcal{Y} \) that maximizes its expected utility:
\[
y^*(m) \in \arg\max_{y \in \mathcal{Y}} \; \mathbb{E}_{a \sim \mu_R(\cdot \mid m, I_R, y)} \left[ u_R(m, a) \right].
\]

An \emph{LLM-Stackelberg equilibrium} is then defined as a pair \( (x^*, y^*(\cdot)) \in \mathcal{X} \times \mathcal{Y}^\mathcal{M} \), such that:
\begin{align*}
x^* &\in \arg\max_{x \in \mathcal{X}} \; \mathbb{E}_{m \sim \mu_S(\cdot \mid I_S, x)}\left[
\mathbb{E}_{a \sim \mu_R(\cdot \mid m, I_R, y^*(m))} \left[ u_S(m, a) \right]
\right], \\
y^*(m) &\in \arg\max_{y \in \mathcal{Y}} \; \mathbb{E}_{a \sim \mu_R(\cdot \mid m, I_R, y)} \left[ u_R(m, a) \right], \quad \forall m.
\end{align*}

This equilibrium captures rational behavior not only at the level of observable strategies but also in the internal reasoning processes that generate those behaviors. Prompts \( x^* \) and \( y^*(\cdot) \) act as meta-strategies that encode high-level planning, recursive belief modeling, and adaptive interpretation. Through the lens of LLM architectures, these prompts structure the agents' generative cognition.

Moreover, this framework enables higher-order reasoning. The sender, when selecting the prompt \( x \), anticipates how the receiver will interpret its generated message \( m \), which also constructs a prompt \( y \) to simulate or infer the latent intent of the sender. This recursive nesting of beliefs, reasoning about reasoning, is operationalized through advanced prompt engineering, chain-of-thought conditioning, and language-based theory of mind capabilities inherent in LLMs.

Unlike classical Stackelberg equilibria which operate over static strategy profiles, the LLM-augmented Stackelberg model is inherently dynamic. Agents can update their prompt strategies over repeated interactions via reinforcement learning, imitation learning, or prompt-tuning based on past performance. Such adaptability is particularly critical in adversarial settings such as cybersecurity, where deception, misaligned incentives, and epistemic uncertainty are prevalent. Here, the reasoning space becomes the battleground for shaping, interpreting, and responding to strategic messages.

The LLM-Stackelberg Stackelberg game extends classical models of strategic communication by embedding LLM-based generative reasoning into both message formation and response selection. The equilibrium concept reflects alignment across both behavioral outcomes and epistemic intentions, enabling cognitively rich interactions in uncertain and adversarial domains. This framework lays a principled foundation for designing agentic AI systems that communicate, adapt, and compete through language-conditioned strategy generation.

\subsection{Prompt Engineering and LLM Tuning}

The construction of agent strategies in the LLM-augmented Stackelberg signaling game relies on how the agents generate and interpret messages, mechanisms that are, in turn, shaped by the behavior of the underlying language models. Two principal methods for influencing LLM behavior are \emph{prompt engineering} and \emph{fine-tuning}, each providing distinct advantages for encoding strategic reasoning, adapting to context, and improving decision quality.

\subsubsection{Prompt Engineering.}
Prompt engineering is the design and control of textual inputs to steer the generative behavior of  LLMs. In the context of the LLM-augmented Stackelberg signaling game, prompts function as meta-strategic variables that influence how agents condition their reasoning and decision policies on observed information.

Formally, let \( x \in \mathcal{X} \) denote the sender's reasoning-level prompt and \( y \in \mathcal{Y} \) the receiver's. The sender observes private information \( I_S \in \mathcal{I}_S \) and selects a prompt \( x = \pi_S(I_S) \), where \( \pi_S : \mathcal{I}_S \to \mathcal{X} \) is a reasoning policy. The LLM then induces a message distribution:
\[
\mu_S(m \mid I_S, x) = \mathbb{P}_{\texttt{LLM}_S}[m \mid x(I_S)],
\]
where \( x(I_S) \) is the realized prompt constructed using observations \( I_S \). Similarly, the receiver observes \( m \) and private context \( I_R \in \mathcal{I}_R \), and constructs a prompt \( y = \pi_R(m, I_R) \), yielding a conditional action distribution:
\[
\mu_R(a \mid m, I_R, y) = \mathbb{P}_{\texttt{LLM}_R}[a \mid y(m, I_R)].
\]

We may generalize prompt engineering as the design of a function class \( \Pi : \mathcal{I} \to \mathcal{X} \) or \( \mathcal{I} \times \mathcal{M} \to \mathcal{Y} \), whose output acts as cognitive scaffolds for reasoning and decision-making. In this framework, the strategic behavior of the agents is shaped by the composition:
\[
I_S \xrightarrow{\pi_S} x \xrightarrow{\texttt{LLM}_S} m, \qquad 
(m, I_R) \xrightarrow{\pi_R} y \xrightarrow{\texttt{LLM}_R} a.
\]

Prompt functions \( \pi_S \) and \( \pi_R \) can be manually specified, learned through meta-optimization, or sampled from a latent prompt distribution \( p(x \mid I_S) \), thus capturing uncertainty or exploration over reasoning modes.

A prompt \( x \in \mathcal{X} \) is typically instantiated as a structured natural language string, such as:
\begin{align*}
x &= \text{``Given the user's past behavior and the intent to appear urgent yet plausible, write a message that maximizes the likelihood of a response.''} \\
y &= \text{``Interpret this message in light of known phishing tactics. Decide whether to click or discard.''}
\end{align*}
These prompts encode procedural knowledge, epistemic models, or belief hierarchies. For example, \( x \) can implicitly simulate a belief distribution over the receiver's expected response policy \( \mu_R(\cdot \mid m, I_R, y) \), enabling higher-order planning.

\subsubsection*{Prompt Classification}

 The prompts used to steer these LLMs can be classified according to their \emph{semantic function}, that is, the kind of cognitive or reasoning structure they activate. This classification helps in designing prompts that align with specific tasks or reasoning styles.

Let \( \mathsf{P} \) denote the space of all semantic prompt types. Within this space, we define four canonical subclasses, each corresponding to a distinct reasoning function:
\begin{align*}
\mathsf{P}_{\text{CoT}} &:\quad \textbf{Chain-of-thought prompts}, \text{ which guide the LLM to perform multi-step reasoning or generate structured explanations}, \\
\mathsf{P}_{\text{Bias}} &:\quad \textbf{Framing and affective prompts}, \text{ which shape the tone, sentiment, or strategic stance of the output}, \\
\mathsf{P}_{\text{ToM}} &:\quad \textbf{Theory-of-mind prompts}, \text{ used to model the beliefs and intentions of other agents}, \\
\mathsf{P}_{\text{Memory}} &:\quad \textbf{Memory prompts}, \text{ designed to preserve context and maintain coherence across multi-turn interactions}.
\end{align*}

Each prompt used by an agent can be decomposed into these semantic components. That is, for a prompt
\[
x \in \mathcal{X} \subset \mathsf{P}_{\text{CoT}} \times \mathsf{P}_{\text{Bias}} \times \mathsf{P}_{\text{ToM}} \times \mathsf{P}_{\text{Memory}},
\]
we can write:
\[
x = \left(x^{\text{CoT}}, x^{\text{Bias}}, x^{\text{ToM}}, x^{\text{Memory}}\right),
\]
where each component \( x^{\cdot} \) activates a specific reasoning module in the LLM. This modularity enables composable and interpretable design of prompt-based policies.

Prompt composition is not one-sided. The receiver agent's prompt \( y \in \mathcal{Y} \) can also be structured similarly, enabling \emph{reflexive cognition}—where both agents simulate, anticipate, and adapt to each other's reasoning pathways. Prompt-based control induces a generative policy:
\[
\mu_S(m \mid I_S, x) = \mathbb{P}_{\texttt{LLM}}(m \mid I_S, x),
\]
where \( I_S \) denotes the agent’s internal state or context, and \( m \) is the message or action sampled from the LLM. The LLM acts as a policy engine, and the prompt \( x \) serves as a control interface. However, this policy is constrained by the LLM’s pretraining distribution \( \mathbb{P}_{\texttt{pretrain}} \), which governs how well prompts align with intended reasoning behaviors.

\paragraph{Geometry and Stability of Prompt Space}

The space \( \mathcal{X} \) of prompts is high-dimensional and non-Euclidean. Small textual changes in prompts can lead to large changes in behavior, a phenomenon called \emph{prompt instability}. To study robustness, we define a distance metric
\[
d : \mathcal{X} \times \mathcal{X} \rightarrow \mathbb{R}_{\geq 0},
\]
and seek the continuity condition:
\[
d(x, x') \ll 1 \quad \Rightarrow \quad \|\mu_S(\cdot \mid I_S, x) - \mu_S(\cdot \mid I_S, x')\|_1 \ll 1.
\]
This expresses the desirable property that similar prompts should induce similar output distributions from the LLM.

\paragraph{Retrieval-Augmented Generation (RAG).}

Retrieval-Augmented Generation (RAG) introduces an external, queryable memory component into the decision-making pipeline, allowing LLMs to dynamically access relevant knowledge during inference. Rather than encoding all domain-specific or strategic information directly into the prompt \( x \) or the LLM parameters \( \phi \), the agent performs a retrieval operation based on its private context \( I_S \) and reasoning intent \( x \).

Formally, let \( \mathcal{K} \) denote a structured external knowledge base or memory store, indexed by a retrieval operator:
\[
\mathcal{R}: \mathcal{X} \times \mathcal{I}_S \rightarrow 2^{\mathcal{K}}, \quad \text{where} \quad \mathcal{R}(x, I_S) = \{k_1, \dots, k_r\}.
\]
Each retrieved item \( k_i \in \mathcal{K} \) may correspond to a prior interaction, adversary tactic, threat indicator, or game-theoretic precedent relevant to the current decision context. The sender's LLM then generates a message via the conditioned distribution:
\[
\mu_S(m \mid I_S, x, \mathcal{R}(x, I_S)) = \mathbb{P}_{\texttt{LLM}}(m \mid I_S, x, \mathcal{R}(x, I_S)),
\]
where the prompt \( x \) and retrieved context jointly shape the generative output. In Stackelberg signaling games, RAG enhances the sender’s strategic expressiveness by enabling \emph{just-in-time reasoning}. For instance, a cyber defense agent might retrieve logs of recent intrusion patterns, policy updates, or incident response precedents, and use this information to craft an informative or deceptive message. In contrast, an attacker could retrieve known vulnerabilities, target behavior profiles, or failed previous attacks to tailor more effective exploits.

RAG also supports temporal coherence and adaptive play in repeated games. The knowledge base \( \mathcal{K} \) can be updated online, enabling agents to learn from prior rounds and refine their retrieval strategies. This facilitates \emph{dynamic situational awareness}, as agents condition behavior on evolving environmental cues rather than relying on static priors or retraining.



\subsubsection{Fine-Tuning LLMs for Strategic Agent Behavior}

LLM, pretrained on general-purpose corpora, can be further adapted to specialized domains or strategic tasks through a process known as \emph{fine-tuning}. This involves updating the model parameters via supervised or reinforcement learning using a task-specific dataset. Unlike \emph{prompt engineering}, which influences model output at inference time without modifying internal weights, fine-tuning embeds long-term behavioral priors and structural patterns into the latent space of the LLM.

Let \( \mathbb{P}_{\phi_0} \) denote a pretrained LLM parameterized by \( \phi_0 \), which defines a conditional distribution over outputs given textual input. Given a labeled dataset
\[
\mathcal{D}_{\text{FT}} = \{(u^{(i)}, v^{(i)})\}_{i=1}^N,
\]
where \( u^{(i)} \in \mathcal{U} \) is a prompt or input, and \( v^{(i)} \in \mathcal{V} \) is the desired output, the goal of fine-tuning is to learn new parameters \( \phi^* \) that minimize the empirical negative log-likelihood:
\[
\phi^* = \arg\min_{\phi} \frac{1}{N} \sum_{i=1}^N -\log \mathbb{P}_\phi(v^{(i)} \mid u^{(i)}).
\]

\paragraph{Game-Theoretic Framing of Preference-Based Fine-Tuning.}
Preference-based fine-tuning has emerged as a promising framework for aligning LLM behavior with human feedback or strategic objectives~\cite{munos2023nash,ye2024online,ye2024theoretical}. Inspired by game theory, these settings are naturally cast as multi-agent games in which each agent’s objective interacts with others’ preferences or policies. Self-play algorithms~\cite{rosset2024direct,ethayarajh2024kto,wu2024self,ye2024evolving,ma2024coevolving,chen2024self,swamy2024minimaximalistapproachreinforcementlearning} instantiate this idea by iteratively improving policies against simulated opponents. The necessity of such game-theoretic solutions is further motivated by the observation of non-transitive preference structures and Condorcet cycles in reward-based models~\cite{liu2025statistical}, where traditional ranking-based alignment breaks down.

Beyond technical alignment, economic models have begun analyzing the incentives behind fine-tuning. Laufer et al.~\cite{laufer2024fine} studied the utility of different stakeholders in LLM adaptation workflows, while Sun et al.~\cite{sun2024mechanism} approached preference aggregation from a mechanism design perspective, proposing Pareto equilibrium as a coordination principle across heterogeneous preference models~\cite{zhong2024panacea}. Other concepts like Bayesian persuasion~\cite{bai2024efficient} and its verbalized form in LLM prompting~\cite{li2025verbalized} further illustrate how agentic models can be designed to influence and align.

\paragraph{Red Teaming via Variational Inference.}
In adversarial settings, fine-tuning takes a strategic form. We consider a two-player game between an attacker (leader) and a defender (follower). The attacker LLM is fine-tuned to simulate threat behaviors using a variational latent variable model. Let \( \mu_\phi \) denote the attacker policy and \( \nu_\psi \) the variational posterior over latent attacker types \( z \in \mathcal{Z} \). Given offline sequences
\[
\mathcal{D}_{\text{att}} = \left\{ (a^{(i)}_{1:T}, o^{(i)}_{1:T}) \right\}_{i=1}^N,
\]
the red-teaming objective becomes maximizing the ELBO:
\[
\mathcal{L}_{\text{ELBO}}(\phi, \psi) = \mathbb{E}_{z \sim \nu_\psi} \left[ \sum_{t=1}^T \log \mu_\phi(a_t \mid a_{<t}, o_{\le t}, z) \right] - \mathrm{KL}\left( \nu_\psi(z \mid a_{1:T}, o_{1:T}) \,\|\, p(z) \right),
\]
optimized via
\[
\max_{\phi, \psi} \sum_{i=1}^N \mathcal{L}^{(i)}_{\text{ELBO}}(\phi, \psi).
\]

\paragraph{Blue Teaming via Preference Alignment.}
The defender LLM is trained to counter adversarial behavior based on inferred preferences. Given interaction context \( c_t \) and preference pairs \( (d_t, d_t') \), the preference model
\[
\mathrm{Pref}_t(d_t, d_t' \mid s_t, a_t, c_t) = \mathbb{P}(d_t \succ d_t')
\]
is used to train a fine-tuned defender policy via Direct Preference Optimization (DPO):
\[
\ell_t(\theta) = -\log \sigma\left( \log \pi_\theta(d_t \mid c_t) - \log \pi_\theta(d_t' \mid c_t) \right) + \lambda \, \mathrm{KL}\left( \pi_\theta(\cdot \mid c_t) \,\|\, \pi_{\text{ref}}(\cdot \mid c_t) \right),
\]
with dynamics governed by:
\[
s_{t+1} \sim P(s_{t+1} \mid s_t, a_t, d_t), \qquad a_t \sim \mu_\phi(\cdot \mid o_{\le t}, a_{<t}).
\]

\paragraph{Equilibrium Policies and Solution Concepts.}
Together, attacker and defender policies form a strategic game over distributions. The fine-tuned pair \( (\mu^*_{\phi}, \mu^*_{\theta}) \) can be interpreted as a Nash equilibrium under partial information and evolving strategies~\cite{munos2023nash}. In multi-agent preference settings, no single output may dominate; thus, alternative solution concepts such as \emph{coarse correlated equilibria}, \emph{evolutionarily stable strategies}, or \emph{Pareto frontiers} are needed to capture the multi-objective nature of alignment~\cite{ye2024theoretical,zhong2024panacea}.

\paragraph{Hybrid Prompt-Finetuned Models.}
While fine-tuning embeds long-term structure, hybrid models enhance adaptivity. Given prompt encoders \( \pi(\cdot) \) and \( \pi'(\cdot) \), the attacker and defender policies are realized as:
\[
\mu_{\text{att}}(a \mid o; \phi^*) = \mathbb{P}_{\phi^*}(a \mid \pi(o)), \qquad 
\mu_{\text{def}}(d \mid c; \theta^*) = \mathbb{P}_{\theta^*}(d \mid \pi'(c)).
\]
This hybrid architecture supports cognitively aligned reasoning and compositional robustness.

\section{LLM Multi-Agent Systems and Games}
LLMs have recently become foundational tools in natural language processing, demonstrating remarkable performance across a wide variety of tasks including translation, summarization, question answering, and reasoning. While the power of individual LLMs continues to grow, an emerging and increasingly important direction is the construction of LLM-based Multi-Agent Systems (MAS), wherein multiple LLMs are deployed as interacting agents. Each agent is configured with a specific role, objective, or expertise, and the collective behavior of the system is shaped by their interactions. This paradigm shift from monolithic, single-agent architectures to distributed, multi-agent frameworks introduces new opportunities and challenges in system design, coordination, and control.

Recent advances exemplify this shift. MAGIS \cite{NEURIPS2024_5d1f0213} demonstrates a specialized multi-agent framework where LLMs assume the roles of developer, QA engineer, and repository custodian to collaboratively resolve GitHub issues. AgentVerse \cite{chen2024agentverse} provides a simulation platform for studying emergent behaviors like cooperation and deception in LLM societies. CAMEL \cite{NEURIPS2023_a3621ee9} explores structured role-playing among agents, revealing dynamics such as negotiation and trust formation. ChatDev \cite{qian2024chatdev} simulates a full software development pipeline through agents with roles like CEO, CTO, and developer. In the medical domain, MedAgents \cite{tang2024medagents} and MDAgents \cite{NEURIPS2024_90d1fc07} showcase how domain-specialized LLM agents can perform diagnostic reasoning and collaborate adaptively. Chain of Agents \cite{NEURIPS2024_ee71a4b1} and ReConcile \cite{chen2024reconcile} highlight communication protocols for long-context processing and consensus-building, respectively, while multi-agent debate \cite{du2023improving} improves factuality via adversarial dialogue.

An LLM-based MAS is characterized by its use of multiple LLM instances as autonomous agents that communicate via natural language. Each agent may operate independently on different aspects of a larger task, and their coordination is typically achieved through structured dialogue protocols, shared memory spaces, or hierarchical task decomposition. This setup mirrors the way human teams collaborate: by dividing labor, exchanging information, and refining their understanding of complex problems through dialogue. Crucially, these systems exploit the inherent language capabilities of LLMs not just for task completion, but also for inter-agent communication, negotiation, and consensus formation.

A central concept in these systems is the multi-agent workflow: a structured pipeline or network of agents that collaboratively execute complex tasks by passing intermediate products, signals, or decisions along well-defined trajectories. Each agent plays a specialized role in the workflow, and the success of the entire system depends on coordinated progress across the workflow stages. In cybersecurity, for instance, an intrusion detection workflow may involve a sequence of LLM agents: the first agent parses system logs for anomalies, a second agent contextualizes anomalies using threat intelligence feeds, a third agent assesses the potential risk, and a fourth agent drafts a response or mitigation plan. Each stage contributes a piece of the larger reasoning chain, and failures or inconsistencies are managed through inter-agent validation and correction.

\subsection{Categories of Multi-Agent Workflows}

To design effective LLM-based multi-agent systems, it is essential to understand the underlying \textit{workflow architectures} that govern inter-agent coordination. These architectures define how tasks are decomposed, how information flows between agents, and how decisions are synthesized. Each type of workflow embodies a different structure of reasoning and communication, and choosing the appropriate one directly influences the system’s performance, robustness, scalability, and interpretability. Particularly in high-stakes domains like cybersecurity, where operations must be timely, resilient, and auditable, workflow architecture plays a pivotal role in shaping both capabilities and limitations.

We now explore four foundational categories of LLM-MAS workflows: \textit{chain}, \textit{star}, \textit{parallel}, and \textit{feedback} workflows. Each corresponds to a different paradigm for organizing agent behaviors and information exchange and is naturally suited to particular cybersecurity applications.

\subsubsection{Chain Workflows}

\begin{figure}[h!]
\centering
\begin{tikzpicture}[node distance=2.2cm, auto]

\tikzstyle{agent} = [inner sep=0pt]

\node[agent] (A1) {\includegraphics[width=1.2cm]{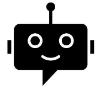}};
\node[agent, right=of A1] (A2) {\includegraphics[width=1.2cm]{robot}};
\node[agent, right=of A2] (A3) {\includegraphics[width=1.2cm]{robot}};
\node[agent, right=of A3] (A4) {\includegraphics[width=1.2cm]{robot}};

\draw[->, thick] (A1) -- (A2);
\draw[->, thick] (A2) -- (A3);
\draw[->, thick] (A3) -- (A4);

\node[below=0.3cm of A1] {Task 1};
\node[below=0.3cm of A2] {Task 2};
\node[below=0.3cm of A3] {Task 3};
\node[below=0.3cm of A4] {Task 4};

\end{tikzpicture}
\caption{Chain workflow: a stepwise pipeline of LLM agents in a linear processing sequence.}
\label{fig:chain_workflow}
\end{figure}

\noindent
Figure~\ref{fig:chain_workflow} depicts a \textbf{chain workflow}, the most intuitive structure for multi-agent coordination. It consists of a linear pipeline of LLM agents, each performing a specific function and passing its output to the next. This assembly-line logic ensures unidirectional information flow and strictly ordered execution, making it well-suited for tasks such as \textit{incident forensics} and \textit{post-breach analysis}. For example, Task 1 may normalize log data from a compromised host, Task 2 flags anomalies, Task 3 cross-references findings with threat intelligence, and Task 4 generates a summary report.

Chain workflows offer high interpretability; each agent's contribution is clearly delineated, aiding debugging and auditing. However, they are prone to \textit{error propagation}: a misclassification early in the sequence can distort downstream reasoning. To mitigate this, resilient designs may include verification agents, rollback mechanisms, or confidence thresholds to contain and correct cascading errors.

\subsubsection{Star Workflows}
\begin{figure}[h!]
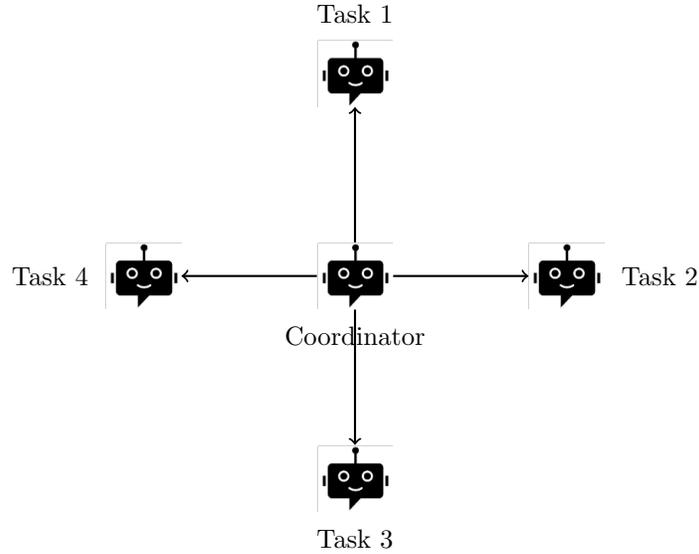

\centering
\begin{tikzpicture}[node distance=1.8cm and 1.8cm, auto]

\tikzstyle{agent} = [inner sep=0pt]

\node[agent] (C) {\includegraphics[width=1.0cm]{robot}};
\node[below=0.1cm of C] {Coordinator};

\node[agent, above=of C] (T1) {\includegraphics[width=1.0cm]{robot}};
\node[above=0.1cm of T1] {Task 1};

\node[agent, right=of C] (T2) {\includegraphics[width=1.0cm]{robot}};
\node[right=0.1cm of T2] {Task 2};

\node[agent, below=of C] (T3) {\includegraphics[width=1.0cm]{robot}};
\node[below=0.1cm of T3] {Task 3};

\node[agent, left=of C] (T4) {\includegraphics[width=1.0cm]{robot}};
\node[left=0.1cm of T4] {Task 4};

\draw[->, thick] (C) -- (T1);
\draw[->, thick] (C) -- (T2);
\draw[->, thick] (C) -- (T3);
\draw[->, thick] (C) -- (T4);

\end{tikzpicture}
\caption{Star workflow: a central LLM agent coordinates specialized analyses across peripheral agents.}
\label{fig:star_workflow}
\end{figure}

\noindent
Figure~\ref{fig:star_workflow} illustrates a \textbf{star workflow}, in which a central LLM agent coordinates specialized peripheral agents, each responsible for an independent analysis. In cybersecurity, this pattern is common in \textit{alert triage} or \textit{risk scoring} scenarios. When a SIEM system detects an anomaly, the coordinator may assign tasks such as behavioral profiling, network forensics, configuration auditing, and compliance checking to distributed agents. Each agent evaluates its dimension in parallel, returning findings to the coordinator for synthesis.

This workflow enables \textit{multi-perspective analysis} and fault-tolerant decision-making. If one agent fails or produces noisy output, the central node can still derive robust judgments from the remaining inputs. However, this architecture requires the coordinator to handle evidence fusion, resolve conflicting inputs, and apply consistent escalation logic, which requires strong \textit{meta-reasoning} capabilities.

\subsubsection{Parallel Workflows}

\begin{figure}[h!]
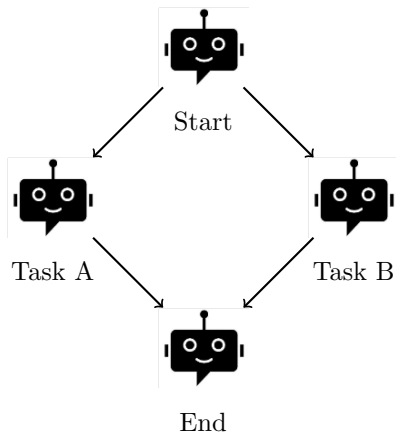

\centering
\begin{tikzpicture}[auto]

\tikzstyle{agent} = [inner sep=0pt]

\node[agent] (Start) at (0,0) {\includegraphics[width=1.2cm]{robot}};
\node[below=0.2cm of Start] {Start};

\node[agent] (A1) at (-2,-2) {\includegraphics[width=1.2cm]{robot}};
\node[below=0.2cm of A1] {Task A};

\node[agent] (A2) at (2,-2) {\includegraphics[width=1.2cm]{robot}};
\node[below=0.2cm of A2] {Task B};

\node[agent] (End) at (0,-4) {\includegraphics[width=1.2cm]{robot}};
\node[below=0.2cm of End] {End};

\draw[->, thick] (Start) -- (A1);
\draw[->, thick] (Start) -- (A2);
\draw[->, thick] (A1) -- (End);
\draw[->, thick] (A2) -- (End);

\end{tikzpicture}
\caption{Parallel workflow: independent LLM agents process distributed data streams before aggregating results.}\label{fig:parallel_workflow}
\end{figure}

\noindent
Figure \ref{fig:parallel_workflow} illustrates a \textbf{parallel workflow} commonly used in \textit{distributed threat hunting}. An initiating agent dispatches LLM-based monitors, Task A and Task B, to operate independently across separate network zones (for example, cloud vs. endpoint logs). Each analyzes telemetry in parallel and reports to a central agent, which consolidates insights into a unified threat assessment.

Parallel workflows are \textit{scalable and latency-efficient}, ideal for environments with large, decentralized data. However, they require \textit{synchronization protocols} to handle inconsistencies and ensure that agents operate under consistent policies, especially when detecting coordinated or stealthy attacks that span multiple domains.




\subsubsection{Feedback (Recurrent) Workflows}
\begin{figure}[h!]
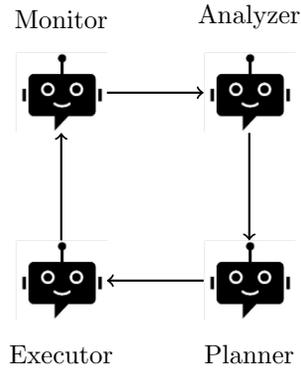

\centering
\begin{tikzpicture}[auto]

\tikzstyle{agent} = [inner sep=0pt]

\node[agent] (A1) at (0, 2.5) {\includegraphics[width=1.2cm]{robot}};
\node[above=0.2cm of A1] {Monitor};

\node[agent] (A2) at (2.5, 2.5) {\includegraphics[width=1.2cm]{robot}};
\node[above=0.2cm of A2] {Analyzer};

\node[agent] (A3) at (2.5, 0) {\includegraphics[width=1.2cm]{robot}};
\node[below=0.2cm of A3] {Planner};

\node[agent] (A4) at (0, 0) {\includegraphics[width=1.2cm]{robot}};
\node[below=0.2cm of A4] {Executor};

\draw[->, thick] (A1) -- (A2);
\draw[->, thick] (A2) -- (A3);
\draw[->, thick] (A3) -- (A4);
\draw[->, thick] (A4) -- (A1);

\end{tikzpicture}
\caption{Feedback workflow: cyclic coordination among LLM agents supporting adaptive cybersecurity operations.}\label{fig:feedback_workflow}
\end{figure}

\noindent
Figure \ref{fig:feedback_workflow} illustrates a \textbf{feedback workflow} typical in adaptive cyber defense. LLM agents collaborate in a closed loop: a \textit{Monitor} detects anomalies, an \textit{Analyzer} interprets threat signals, a \textit{Planner} formulates responses, and an \textit{Executor} enacts mitigations, cycling back to assess the impact of interventions.

Such workflows are critical for \textit{active defense}, \textit{red-vs-blue simulations}, and \textit{cyber deception}, where agents must iteratively adapt to adversarial tactics. By enabling real-time strategic refinement, feedback workflows unlock resilient and intelligent operations. However, they must be carefully engineered with safeguards, such as bounded iterations or convergence thresholds, to avoid instability and ensure mission coherence.

\subsubsection{Heterogenous Workflow}

Many real-world systems employ heterogeneous workflows that combine LLM agents with human operators, leveraging the speed and scalability of AI alongside human judgment, ethics, and domain expertise. In these human-AI teaming architectures, humans may guide policy, supervise execution, or audit outcomes, while LLMs act as assistants, co-pilots, or adversarial simulators depending on the task.

Cybersecurity incident response is a prime example. Upon alert, LLM agents parse logs, extract indicators, and propose mitigation steps. Human analysts then validate outputs, assess risks, and make final decisions, ensuring oversight where full automation is inappropriate.

These workflows also support interactive feedback loops, where agents propose strategies (e.g., for patching or risk assessment), and humans iteratively refine or constrain them. The LLM incorporates feedback to generate updated recommendations, enabling a co-adaptive decision process that blends automation with expert input.

\subsubsection{Hybrid Workflows}

\begin{figure}[h!]
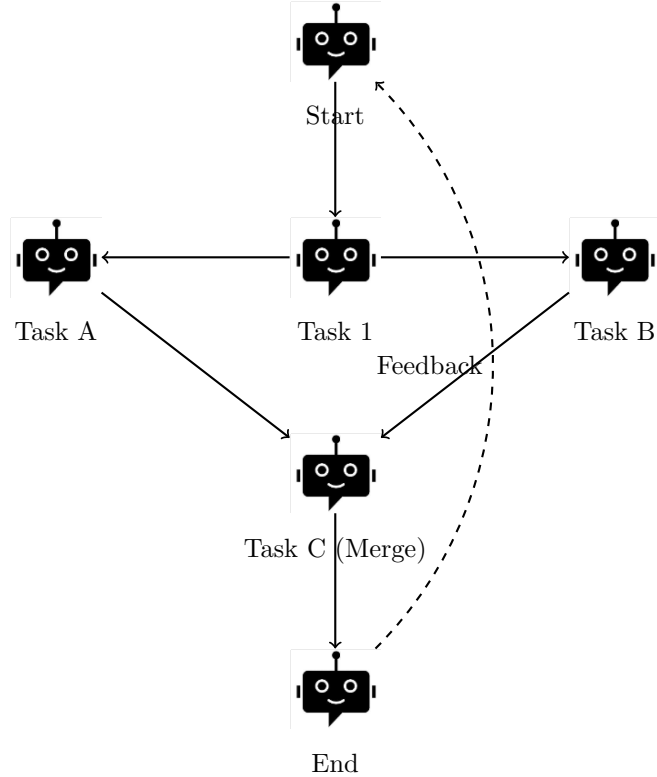

\centering
\begin{tikzpicture}[auto, node distance=1.8cm and 2.5cm]

\tikzstyle{agent} = [inner sep=0pt]

\node[agent] (Start) {\includegraphics[width=1.2cm]{robot}};
\node[below=0.2cm of Start] {Start};

\node[agent, below=1.8cm of Start] (T1) {\includegraphics[width=1.2cm]{robot}};
\node[below=0.2cm of T1] {Task 1};

\node[agent, left=2.5cm of T1] (A1) {\includegraphics[width=1.2cm]{robot}};
\node[below=0.2cm of A1] {Task A};

\node[agent, right=2.5cm of T1] (A2) {\includegraphics[width=1.2cm]{robot}};
\node[below=0.2cm of A2] {Task B};

\node[agent, below=1.8cm of T1] (Recon) {\includegraphics[width=1.2cm]{robot}};
\node[below=0.2cm of Recon] {Task C (Merge)};

\node[agent, below=1.8cm of Recon] (End) {\includegraphics[width=1.2cm]{robot}};
\node[below=0.2cm of End] {End};

\path[->, thick, dashed, bend right=45] (End) edge node[left] {Feedback} (Start);

\draw[->, thick] (Start) -- (T1);
\draw[->, thick] (T1) -- (A1);
\draw[->, thick] (T1) -- (A2);
\draw[->, thick] (A1) -- (Recon);
\draw[->, thick] (A2) -- (Recon);
\draw[->, thick] (Recon) -- (End);

\end{tikzpicture}
\caption{Hybrid workflow with a sequential initialization, parallel branching, and a convergence task with feedback to the initial stage.}
\end{figure}\label{fig:hybrid_workflow}

Real-world multi-agent systems rarely follow a single architectural pattern. Instead, they adopt hybrid workflows that blend parallel, sequential, and feedback structures to optimize modularity, efficiency, and resilience. For instance, an LLM-based incident response system might first use a parallel layer to collect evidence from distributed sensors, then a star configuration for centralized triage, followed by a feedback loop between human analysts and LLM agents to refine mitigation strategies.

The depicted workflow in Figure \ref{fig:hybrid_workflow} exemplifies such a design. It begins with a Start node, triggered by an event like a suspicious login, which launches Task 1 for telemetry parsing and threat contextualization. The system then branches into Task A, which performs anomaly detection, and Task B, which conducts threat intelligence enrichment. These parallel analyses are fused in Task C, an aggregation module that scores threats and recommends actions. The End node executes mitigation or generates a report, while a feedback loop enables policy refinement and prompt adaptation, ensuring continual learning and response optimization.

\bigskip

\subsection{Robustness and Resilience of the MAS Workflow}

One of the core motivations for adopting multi-agent architectures in LLM-based systems is the need to mitigate inherent weaknesses of individual models. Large Language Models, despite their fluency and versatility, are susceptible to issues such as \textit{hallucination}, \textit{context forgetting}, \textit{inconsistency}, and susceptibility to adversarial prompting. When operating in isolation, an LLM may produce outputs that are syntactically plausible but semantically incorrect or operationally dangerous—an unacceptable risk in high-stakes domains like cybersecurity. To address these challenges, MAS workflows are designed to enhance \textbf{robustness} and \textbf{resilience} through architectural redundancy, inter-agent verification, and error isolation mechanisms.

\textbf{Robustness} in this context refers to the system’s ability to maintain accuracy and reliability despite noisy inputs, partial failures, or unpredictable model behavior \cite{zhu2024disentangling}. Multi-agent workflows achieve this by deploying multiple agents to perform overlapping or complementary analyses. For example, in a vulnerability assessment workflow, several LLM agents may independently evaluate the exploitability of a newly reported vulnerability. These agents might differ in their prompting strategies, external data access, or prior knowledge bases. If the agents converge on a consistent rating, confidence in the output increases. If they disagree, a supervisory or meta-agent may invoke a reconciliation protocol, such as weighted voting, argument-based debate, or external validation via trusted data sources. This layered redundancy ensures that no single agent’s error can dominate the system’s behavior, significantly improving fault tolerance.

\textbf{Resilience}, on the other hand, refers to the system’s capacity to recover from disruptions, adapt to new threats, and sustain critical operations in the face of failures \cite{zhu2024foundations,kott2019cyber}. MAS workflows enhance resilience by supporting dynamic reconfiguration and adaptive routing. In a star workflow, for instance, if a peripheral agent becomes unresponsive or compromised, the central agent can reroute tasks to a backup agent or reassign roles dynamically. In feedback workflows, agents can revise earlier decisions based on new information from downstream processing, enabling iterative correction and long-term learning. These capabilities are crucial in cybersecurity scenarios such as real-time intrusion detection, where novel attack vectors may trigger cascading inconsistencies that require immediate containment and reasoning recalibration.

Furthermore, workflow-level resilience is supported by mechanisms for memory persistence, behavior logging, and provenance tracking. These features allow systems to not only correct errors but also \textit{learn from them}. For instance, if a phishing detection agent incorrectly classifies a benign email as malicious, the system can trace the reasoning path, identify where hallucination or overfitting occurred, and adjust future prompts or priors accordingly. Over time, this adaptive feedback loop reinforces both the structural robustness of the workflow and the epistemic resilience of the agents.

\subsection{Optimization and Coordination of the Workflow}

Beyond robustness and resilience, the effectiveness of a multi-agent system (MAS) ultimately hinges on the efficiency and coherence of agent interactions. \textbf{Coordination protocols} serve as the backbone of workflow optimization, allowing agents to communicate, align goals, resolve disagreements, and make collective decisions. These protocols ensure that the distributed components of the system do not merely function in parallel but operate in concert toward shared objectives. In the context of LLM-based MAS, where communication occurs through natural language and decisions are often non-deterministic, coordination becomes both a design challenge and a rich domain for optimization.

Several classes of coordination protocols have been developed to manage agent interactions. \textit{Voting mechanisms} allow agents to independently propose outputs, with the majority or weighted majority determining the final decision. This approach is particularly useful in security-relevant workflows, such as malware classification, where the aggregation of multiple perspectives can dampen the effects of hallucinations or model bias. Systems like ReConcile \cite{chen2024reconcile} and Chain of Agents \cite{NEURIPS2024_ee71a4b1} exemplify this, combining intermediate reasoning across agents to enhance final output fidelity.

\textit{Debate protocols} go beyond passive aggregation by allowing adversarial reasoning between agents. For instance, \cite{du2023improving} introduces structured debates to reduce hallucinations and improve factuality, even in black-box settings. This is especially powerful in threat attribution or misinformation detection workflows, where subtle inconsistencies require adversarial challenge to be surfaced.

\textit{Reconciliation frameworks} employ tree-like dialogue structures to synthesize divergent agent outputs into a refined consensus. Systems like MedAgents \cite{tang2024medagents} demonstrate structured dialogues among specialists, while CAMEL \cite{NEURIPS2023_a3621ee9} facilitates social reasoning and consensus-building in simulated LLM societies.

From a theoretical standpoint, optimizing MAS workflows involves questions traditionally studied in systems theory, control, and distributed computation. For instance, one must ask: Are agent behaviors \textbf{controllable}? That is, can the system guide agent states toward desired outputs under defined inputs and constraints? Are they \textbf{observable}; can the system infer internal agent states or errors from their external outputs? These properties are crucial to ensure that the MAS can be debugged, supervised, and aligned with human-intended goals.

\textbf{Stability} is another key concern, especially in workflows that involve feedback or recurrent interaction. Without carefully designed dynamics, agent interactions may oscillate, deadlock, or produce non-convergent outputs. Formal techniques from dynamical systems, such as Lyapunov stability analysis or bounded regret learning, may offer tools for analyzing and guaranteeing convergence in MAS workflows.

Furthermore, optimization in LLM-based MAS involves designing protocols for \textbf{resource-aware task allocation}, \textbf{response latency control}, and \textbf{confidence-weighted reasoning}. For example, in ChatDev \cite{qian2024chatdev} and MAGIS \cite{NEURIPS2024_5d1f0213}, tasks are assigned to agents based on their specialization and system phase (e.g., planning vs. testing), enabling more efficient and relevant collaboration. Reinforcement learning or multi-armed bandit frameworks can be used to adaptively select coordination strategies that minimize risk and maximize mission impact.

In cybersecurity and high-stakes domains, such optimization and coordination mechanisms are not theoretical luxuries. They are operational necessities. Automated systems must not only detect and respond to threats quickly, but also behave predictably under stress, degrade gracefully in the face of failure, and allow human operators to understand and influence their decisions. The field therefore demands coordination strategies that are both mathematically grounded and practically effective.

\subsection{Gestalt Game-Theoretic Description}

To characterize the interaction of multi-agent systems (MAS) in a workflow setting, it is essential to model both collaborative and adversarial dynamics. On one hand, agents coordinate to complete tasks and propagate information through shared objectives; on the other hand, adversarial agents may attempt to disrupt or deceive, prompting defensive behaviors from cooperating agents. These interactions span multiple scales, echelons, and temporal layers, requiring a holistic formulation that transcends traditional flat game-theoretic models.

To address this complexity, we adopt the \emph{games-in-games} framework introduced in \cite{chen2019control}, which provides a system-of-systems perspective to model layered decision-making among heterogeneous and possibly adversarial agents. This framework captures the interleaving of cooperative Nash games (e.g., among network operators or workflow agents) and adversarial Stackelberg games (e.g., between defenders and strategic attackers), thereby embedding both horizontal and vertical game structures. The resulting formulation supports a \emph{meta-equilibrium} solution concept that reflects the coupled reasoning and strategic anticipation among multiple decision layers.

This holistic structure gives rise to what we call a view \emph{Gestalt game-theoretic}, a synthesis in which the whole interactional system displays emergent strategic behavior that is not decomposable into isolated pairwise games. The Gestalt perspective aligns with recent developments in coupled game theory (cf. \cite{pawlick2015flip}), where the equilibrium behavior of agents is shaped by interdependencies between signaling, control, and information asymmetries across components. Such games-in-games structures are particularly relevant in MAS workflows subjected to cyber-physical threats (cf. \cite{zhu2015game}), where interactions must be evaluated in the presence of uncertainty, deception, and partial observability.

Examples of Gestalt games include the coordination of mobile autonomous systems under adversarial jamming \citep{chen2019control}, or trust management in cloud-connected cyber-physical devices under APT attacks modeled through coupled signaling and FlipIt games \citep{pawlick2015flip,pawlick2017strategic,pawlick2018istrict}. These formulations have captured how security, coordination, and information trust co-evolve in interconnected agent networks.

\section{Conclusions}

This chapter has outlined a transformative vision at the intersection of game theory, agentic AI, and cybersecurity. We have introduced a new generation of game-theoretic models, Nash and Stackelberg games based on LLM that leverage large language models not only as reasoning engines but also as generative policy mechanisms. These models relax traditional assumptions of perfect rationality, common knowledge, and symbolic optimization, allowing for richer, language-conditioned strategies grounded in prompt-based reasoning.

By extending classical equilibrium concepts to the prompt space, we have shown how strategic behavior can emerge from iterative, context-aware interactions among LLM agents. These developments mark a significant departure from static game formulations, reflecting the cognitive flexibility and epistemic asymmetry of real-world cyber operations.

Furthermore, we have examined the role of LLM-based multi-agent systems  as the architectural foundation for scalable, interpretable, and adaptive security solutions. Through various workflow structures, including chain, star, parallel, and gestalt configurations, we demonstrated how coordinated agent behaviors can produce emergent capabilities in threat detection, deception, red-teaming, and collaborative defense.

The integration of game-theoretic reasoning with LLM-powered agents paves the way for designing AI systems that are not only responsive to dynamic threats, but also capable of modeling intent, simulating adversarial cognition, and adapting through self-refinement. This synergy between theory and generative intelligence introduces a paradigm shift in cybersecurity, where strategies are no longer static prescriptions but dynamically synthesized outcomes of language-based deliberation.

Looking ahead, future research should explore the training and evaluation of LLM agents in high-stakes security environments, the governance of multi-agent reasoning under adversarial pressure, and the development of human-agent teaming protocols that preserve trust and accountability. As agentic AI continues to evolve, its coupling with game-theoretic principles will remain essential for ensuring that intelligent cyber systems are robust, secure, and aligned with human intent.

\bibliographystyle{spbasic}
\bibliography{reference.bib} 
\end{document}